\documentclass[prc]{revtex4}
\usepackage{graphicx}

\begin{document}

\newcommand{\ie}{{\it i.e.}}
\newcommand{\eg}{{\it e.g.}}
\newcommand{\etc}{{\it etc.}}
\newcommand{\cf}{{\it cf.}}
\newcommand{\etal}{{\it et al.}}
\newcommand{\be}{\begin{eqnarray}}
\newcommand{\ee}{\end{eqnarray}}
\newcommand{\jp}{$ J/ \psi $}
\newcommand{\pp}{$ \psi^{ \prime} $}
\newcommand{\ppp}{$ \psi^{ \prime \prime } $}
\newcommand{\dd}[2]{$ #1 \overline #2 $}
\newcommand{\noi}{\noindent}

\title{Bottom Tetraquark Production at RHIC?} 

\author{R. Vogt}
\affiliation
{Nuclear and Chemical Sciences Division,
Lawrence Livermore National Laboratory, Livermore, CA 94551,
USA}
\affiliation
    {Department of Physics and Astronomy,
University of California, Davis, CA 95616,
USA}

\author{A. Angerami}
\affiliation
{Nuclear and Chemical Sciences Division, 
Lawrence Livermore National Laboratory, Livermore, CA 94551,
USA}

\begin{abstract}
{\bf Background:}   A resonance has been observed by the ANDY Collaboration at the Relativistic
  Heavy-Ion Collider at Brookhaven National Laboratory
  in Cu+Au collisions at center-of-mass
  energy $\sqrt{s} = 200$~GeV and at forward rapidity
  with an average mass of 18.15~GeV.  The Collaboration suggests that it is a
  $b \overline b b \overline b$ tetraquark
  state decaying to two $\Upsilon$(1S)
  states, each measured through the $\Upsilon \rightarrow ggg$ channel.
  {\bf Purpose:}   Their suggestion is investigated assuming that the two $\Upsilon$ states
  are produced through the materialization of
  a $|uud b\overline b b \overline b \rangle$ Fock state in the projectile.
  {\bf Methods:} The $\Upsilon$ pair mass and rapidity distributions arising from
  such a state are calculated.  The production of an
  $X_b(b \overline b  b \overline b)$ tetraquark state from the same Fock
  configuration is also investigated.  The dependence on bottom quark mass and their transverse momentum range is also studied.
  {\bf Results:} It is found that double $\Upsilon$ production from these $|uud b \overline b b \overline b \rangle$ states peak in the rapidity range of the ANDY detector.  The $\Upsilon$ pair and $X_b$ masses are, however, higher than the mass reported by the ANDY Collaboration.
  {\bf Conclusions:}   The results obtained from these
  calculations are incompatible with the ANDY result.  They are, however,
  compatible with previous predictions of $b \overline  b b \overline b$
  tetraquark masses.
\end{abstract}

\maketitle

\section{Introduction}
\label{intro}

Quantum Chromodynamics (QCD) is the theory describing the interactions among
quarks and gluons. The
fundamental color charges in this theory are confined in
color-neutral objects known as baryons and mesons, containing three quarks and 
quark-antiquark pairs respectively. Since the advent of QCD, the existence
of other, exotic, hadrons outside the conventional quark model, such as
tetraquarks and pentaquarks, consisting of
four or five valence quarks respectively, have been
postulated, as early as in Murray Gell-Mann's introduction of the quark model
\cite{Gell-Mann:1964ewy}.

Most exotic hadrons so far discovered contain at least one
$c \overline c$ pair.  The first proposed tetraquark candidate,
the $X(3872)$, measured by Belle \cite{Belle:2003nnu} in $e^+ e^-$ collisions,
has been observed in many systems, including in nucleus-nucleus collisions
\cite{CMS:2021znk}.  The $Z(4430)$, a $c \overline c d \overline u$ tetraquark
candidate was first measured by Belle \cite{Belle:2007hrb} and later
confirmed by LHCb \cite{LHCb:2014zfx}.  The $Z(3900)$, reported by BES III
\cite{BESIII:2013ris} and Belle \cite{Belle:2013yex}, was confirmed as a
four-quark state.  Most recently, LHCb announced the discovery of a
$c \overline c c \overline c$ tetraquark, the $X(6900)$ \cite{LHCb:2020bwg}.
LHCb has also announced the discovery of a $uudc \overline c$ pentaquark state,
the $P_c(4312)^+$ \cite{LHCb:2019kea}.  So far no tetraquark states
containing $b$ quarks, either $q \overline q b \overline b$
or $b \overline b b \overline b$, have been confirmed.

The ANDY Collaboration has recently reported an observation of a resonance with
a mass of 18.15~GeV in Cu+Au collisions at the Relativistic Heavy-Ion Collider
(RHIC) \cite{ANDY}. This observation
was made at forward rapidity and was interpreted as a
$b \overline b b \overline b$ tetraquark state decaying to two $\Upsilon$(1S)
states, each decaying to hadrons through the $\Upsilon \rightarrow ggg$ channel.
These data were taken at 
$\sqrt{s_{NN}} = 200$~GeV in 2012.  They used a minimum bias trigger plus an
inclusive jet and dijet trigger.  Because no luminosity measurement was
performed, the results were presented as fractions of the minimum bias yields.  

In that analysis, multiple jets were found in a given
event.  Through an event-mixing study they found that only high energy dijets
exhibited azimuthal angular correlations.  No such correlations were observed
for low energy dijets.  They constructed the dijet mass for these large
energy dijets and found peaks at $M_{\rm dijet} = 17.83 \pm 0.2$~GeV for dijets
with energy $250 < E < 260$~GeV and at
$M_{\rm dijet} = 18.47 \pm 0.22$~GeV for dijets with energy
$260 < E < 270$~GeV.  Both peaks have high statistical significance,
$9\sigma$ and $8.4\sigma$ respectively, since there is little background in
this region.  Combining the two dijet energy bins gives an average dijet mass
of $M_{\rm dijet} = 18.12 \pm 0.15$~GeV.  The two jets comprising the
dijet were measured within the calorimeter acceptance $3.0 < \eta < 3.5$
\cite{ANDY}.
They then looked for candidate $\Upsilon$(1S) decays to three gluons and found
evidence for double $\Upsilon$ production, both of which decayed hadronically
through $\Upsilon({\rm 1S}) \rightarrow 3g$.  They concluded that the best
candidate for their dijet mass signal is a
$X_b(b \overline b b \overline b)$ tetraquark state.

The authors of Ref.~\cite{ANDY} noted that there have been many predictions of
an $X_b$ tetraquark mass, all of them larger than the ANDY value of 18.12~GeV.
Karliner, Rosner and Nussinov \cite{Karliner} obtained a mass of 18.826~GeV
based on meson and baryon mass systematics.  A similar mass value,
$18.84 \pm 0.09$~GeV, was found by Wang \cite{Wang}, based on QCD sum rules.
Calculations of the ground state $X_b$ mass by Bai, Lu and Osborne \cite{Bai}
and Wu {\it et al.} \cite{Wu} found lighter masses of 18.69~GeV and 18.46~GeV
respectively.
Other calculations of $b \overline b b \overline b$ tetraquark states
\cite{GangYang,XZWeng} predict $\Upsilon(1{\rm S})\Upsilon(1{\rm S})$
tetraquark states with
$J^{PC} = 1^{+-}$ with masses of $\approx 19$~GeV, well above the ANDY mass but
compatible with the calculations of Refs.~\cite{Karliner,Wang}.
Lattice QCD calculations \cite{lattice}, on the other hand, found no evidence
for the all $b$ tetraquark while an analysis by Richard, Valcarce and Vijande
\cite{Richard} suggested that such a state wound be unbound.

Experimental searches for $X_b$ tetraquarks have been carried out in $p+p$
collisions at the LHC.
LHCb set limits on the $X_b$ mass through
$\Upsilon ({\rm 1S}) + \Upsilon^\star \rightarrow \mu^+ \mu^- \mu^+ \mu^-$ in
the rapidity range $2 < \eta < 5$ in $p+p$ collisions
at $\sqrt{s} = 7$, 8, and 13~TeV \cite{LHCbXb}.  They found no candidate
events in the mass range $17.5 < M_{X_b} < 20$~GeV.  CMS studied
$\Upsilon ({\rm 1S}) + \Upsilon^\star \rightarrow \mu^+ \mu^- \mu^+ \mu^-$ and
$\mu^+ \mu^- e^+ e^-$ at midrapidity in $p+p$ collisions at $\sqrt{s} = 13$~TeV
\cite{CMSXb}.  They saw no evidence of a signal in the mass range
$17.5 < M_{X_b} < 19.5$~GeV \cite{CMSXb}.

Pair production of quarkonium
has been measured before, most recently at collider
energies.  For example, in the same analysis that set limits on $X_b$ production
at $\sqrt{s} = 13$~TeV, the CMS Collaboration measured double $\Upsilon$(1S)
production through both single and double parton scattering \cite{CMSXb}.

The LHCb Collaboration reported $J/\psi$ pair production originating from
$b \overline b$ pair production, followed by the decay of both $b$ hadrons to
$J/\psi$ \cite{LHCb_bbbar}.  This LHCb measurement, performed in
$p+p$ collisions at $\sqrt{s} =7$ and 8~TeV, at forward rapidity,
$2.5 < y < 5$, can be well described by calculations assuming that the
production originates from a single
$b \overline b$ pair \cite{LHCb_bbbar,RVazi2}.  The predominant $b \overline b$
production mechanism in $p+p$ collisions, perturbative QCD, produces single
$b \overline b$ pairs at $y = 0$.  The LHCb rapidity range is not far forward
relative to the fractional momentum carried by the gluons initiating the
$b \overline b$ production.  

The $\Upsilon$ pair production suggested by ANDY, on the other hand, is in a
kinematic region more similar to fixed-target $J/\psi$ pair production
measured by NA3 with $\pi^-$ beams of laboratory momenta 150 and
280~GeV/$c$
\cite{Badpi} and 400~GeV/$c$ proton beams \cite{Badp}.
The fraction of the $\pi^-$ momentum carried by the $J/\psi$ pair in 
$\pi^-N \rightarrow J/\psi J/\psi X$ events was
$x_{\psi \psi} \geq 0.6$ at 150 GeV/$c$ and $x_{\psi \psi} \geq 0.4$ at
280~GeV/$c$.  In perturbative QCD, $J/\psi$ pair production is near
$x_{\psi \psi} \sim 0$.

The 
average invariant mass of the $J/\psi J/\psi$ pairs measured by NA3 was well
above the $2m_\psi$ threshold while the average transverse
momentum of the pair was small, suggesting tightly correlated
production \cite{Badpi,Badp}.
These measurements were studied within the intrinsic charm model, assuming
production from six and seven-particle Fock states,
$|\overline u d c \overline c c \overline c \rangle$ and
$|uud c \overline c c \overline c \rangle$ respectively, in
Ref.~\cite{dblic}.   Good agreement of the calculations with the measured mass
distributions was found.  Other mechanisms for double
$J/\psi$ production were further studied in Ref.~\cite{dbljpsi}, none of
which resulted in similarly good agreement with the NA3 data.

The possibility of double $\Upsilon$ production from a similar Fock state
would be even more rare assuming intrinsic heavy flavor production scales as
$(m_c^2/m_b^2)^2$ for the production of two  $Q \overline Q$ pairs in the same
Fock state \cite{dblic}.  The higher energies at RHIC allow
production of these more massive states.

This work investigates the ANDY result assuming that it could arise from direct
$\Upsilon$(1S) pair production or from production of an
$X_b(b \overline b b \overline b)$ state in QCD, assuming they arise from
intrinsic heavy quark Fock states \cite{intc1,intc2}.
Both potential production channels are assumed to be produced
from a single Fock state of the nucleon, in particular
$|uud b \overline b b\overline b \rangle$, in a single interaction, {\it i.e.}
double parton scattering is not considered.  Intrinsic heavy quark states are
an especially attractive candidate for the ANDY signal because they are
manifested at large Feynman $x$ instead of at central rapidities, as in the
case of perturbative production.  However, the ANDY detector at RHIC
covers the more forward region in pseudorapidity $3 < \eta < 3.5$, larger
rapidity than covered by previous measurements of heavy flavor at RHIC.  

Here double $\Upsilon$ and $X_b(b \overline b b \overline b)$ production is
calculated through the 
materialization of double intrinsic $b \overline b$ Fock components of the
nucleons.  Note that even though the light quark content of the projectile can
make a difference in which open heavy flavor states are most likely to be
produced \cite{VBH2,RVSJB_asymm,tomg}, quarkonium states such as $J/\psi$ and
$\Upsilon$ share no valence quarks with the colliding beams and would thus
be equally produced by protons
and neutrons, {\it i.e.}, they are not leading particles.

Single $\Upsilon$ production from a
$|uudb \overline b \rangle$ state is first described in Sec.~\ref{single_Ups}.
Next, $\Upsilon$ pair production from a $|uudb \overline b b\overline b \rangle$
state is described in
Sec.~\ref{double_Ups}.  The Feynman-$x$, rapidity and pair mass
distributions are calculated.  The dependence of the mass distributions on the
$b$ quark mass and internal $k_T$ dynamics of the quarks in the
Fock state are shown.  The possibility that the state is manifested instead
as an $X_b$ tetraquark is discussed and these distributions compared and
contrasted to those for double $\Upsilon$ production in Sec.~\ref{Xb_tet7}.
The results are
related to the ANDY kinematics.  Conclusions are drawn 
in Sec.~\ref{summary}.

\section{Intrinsic Heavy Flavor Production}

The wave function of a proton in QCD can be represented as a
superposition of Fock state fluctuations of the $\vert uud \rangle$ state,
{\it e.g.}\ $\vert uudg
\rangle$, $\vert uud q \overline q \rangle$, $\vert uud Q \overline Q \rangle$,
\ldots.
When the projectile scatters in the target, the
coherence of the Fock components is broken and the fluctuations can
hadronize \cite{intc1,intc2,BHMT}.  These
intrinsic $Q
\overline Q$ Fock states are dominated by configurations with
equal rapidity constituents, so that the intrinsic heavy quarks carry a large
fraction of the projectile momentum \cite{intc1,intc2}.  Heavy quark hadrons
can be formed by coalescence, either with light quarks {\it e.g.} to form
a $\Lambda_c^+(udc)$ and $D^0(u \overline c)$ from a $|uudc \overline c \rangle$
state or a final-state $J/\psi$ with a proton.  Leading charm asymmetries
have been measured as a function of $x_F$ and $p_T$
in fixed-target $\pi^- + p$ interactions where a
$D^-(d \overline c)$, which can be produced from a
$|\overline u d c \overline c \rangle$ Fock state of the negative pion, is
leading over a $D^+ (\overline d c)$ which is not \cite{RVSJB_asymm}.
It is worth
noting that both $D^+$ and $D^-$ can be produced at higher $x_F$ than purely
perturbative production in a higher Fock state, namely a six-particle
$|\overline u d d \overline d c \overline c \rangle$ state, but there would be
no difference in their distributions, neither would lead the other when produced
from this state.  In addition, the average $x_F$ for $D$ mesons hadronized from
this state would be lower than the $D^-$ average $x_F$ from the minimal Fock
state required to produce it, the four-particle
$|\overline u d c \overline c \rangle$ state \cite{tomg}.  These higher Fock
states would also have lower probabilities for manifestation from the projectile
hadron.

In this work, the formulation for intrinsic heavy quarks
in the proton wavefunction
postulated by Brodsky and collaborators in Refs.~\cite{intc1,intc2} has been
adapted.  That work was more specifically directed toward charm quarks.
There are also other variants of intrinsic charm distributions in the proton,
including meson-cloud models where the proton fluctuates into a
$\overline D(u \overline c) \Lambda_c (udc)$ state
\cite{Paiva:1996dd,Neubert:1993mb,Steffens:1999hx,Hobbs:2013bia}, also resulting
in forward production, or a
sea-like distribution \cite{Pumplin:2007wg,Nadolsky:2008zw}, only enhancing the
distributions produced by massless parton splitting functions as in DGLAP
evolution.  Intrinsic charm
has also been included in global analyses of the parton densities
\cite{Pumplin:2007wg,Nadolsky:2008zw,Dulat:2013hea,Jimenez-Delgado:2014zga,NNPDF_IC}.  (See Ref.~\cite{Blumlein} for a discussion of a possible
kinematic constraint on intrinsic charm in deep-inelastic scattering.)

The probability of intrinsic charm production, $P_{{\rm ic}\, 5}^0$, obtained
from these analyses, as well as others, has been
suggested to be between 0.1\% and 1\%.  The reviews in 
Refs.~\cite{IC_rev,Stan_review} describe the global analyses and other
applications of intrinsic heavy quark states.  
New evidence for a finite charm quark asymmetry in the
nucleon wavefunction from lattice gauge theory, consistent with intrinsic
charm, was presented in Ref.~\cite{Sufian:2020coz}.  

The general consensus is that the
probability of intrinsic bottom production, $P_{{\rm ib}\, 5}^0$,
will scale as the square of the quark mass,
$m_c^2/m_b^2$, for production from a minimal Fock state configuration such as
$|uud Q \overline Q \rangle$ where $Q = c, b$.
A few calculations of intrinsic bottom production have been made previously
\cite{Bottom_Prod,RV_SJB_ib}.  Some additional prior results are also
summarized in Ref.~\cite{IC_rev}.  

Here single $\Upsilon$ production from such a minimal Fock state configuration
is summarized first with differences between results for charm and bottom
highlighted.  Starting from this baseline, single and double $\Upsilon$
production from the minimal Fock state configuration for $\Upsilon$ pair
production, $|uud b \overline b b \overline b \rangle$, is developed.  The
$x_F$ and rapidity distributions are described and the $\Upsilon$ pair mass
distributions are presented and the sensitivities of these distributions to
calculational inputs are discussed.  Some attention is paid to the normalized
cross section for such states but the main focus of the discussion is whether
the distributions produced in this approach are compatible with the kinematic
range of
the ANDY measurement and, if so, are the resulting mass distributions at all
compatible with their measured mass.

\subsection{Single $\Upsilon$ Production from a $|uud b \overline b \rangle$ State}
\label{single_Ups}

Production of a single
$\Upsilon$ from a five-particle proton Fock state is considered first, analogous
to $J/\psi$ production from such a state, as recently studied in
Ref.~\cite{RV_SeaQuest}.  In the case of $\Upsilon$ production,
the frame-independent probability distribution of a $5$-particle
$b \overline b$ Fock state in the proton is 
\be
dP_{{\rm ib}\, 5} = P_{{\rm ib}\,5}^0
N_5 \int dx_1 \cdots dx_5 \int dk_{x\, 1} \cdots dk_{x \, 5}
\int dk_{y\, 1} \cdots dk_{y \, 5} 
\frac{\delta(1-\sum_{i=1}^5 x_i)\delta(\sum_{i=1}^5 k_{x \, i}) \delta(\sum_{i=1}^5 k_{y \, i})}{(m_p^2 - \sum_{i=1}^5 (m_{T \, i}^2/x_i) )^2} \, \, ,
\label{icdenom5}
\ee
where $i = 1$, 2, 3 are the interchangeable light quarks ($u$, $u$, $d$)
and $i = 4$ and 5 are the $b$ and $\overline b$ quarks respectively.
Here $N_5$ normalizes the
$|uud b \overline b \rangle$ probability to unity and $P_{{\rm ib}\, 5}^0$
scales the unit-normalized
probability to the assumed intrinsic bottom content of the proton.  The delta
functions conserve longitudinal ($z$) and transverse ($x$ and $y$) momentum.
The denominator of Eq.~(\ref{icdenom5}) is
minimized when the heaviest constituents carry the largest fraction of the
longitudinal momentum, $\langle x_b \rangle > \langle x_q \rangle$.  Given
that $m_Q \gg m_q$, one does not expect large differences between
$\langle x_b \rangle$ and $\langle x_c \rangle$ from $|uudb \overline b \rangle$
and $|uudc \overline c \rangle$ states respectively.

In Ref.~\cite{RV_SeaQuest}, the $J/\psi$ $p_T$ distribution from intrinsic
charm was calculated for the first time by integrating over the light and charm
quark $k_T$ ranges in Eq.~(\ref{icdenom5}).  In that work, $k_{T \, q}^{\rm max}$
was set to 0.2~GeV while the default for $k_{T \, c}^{\rm max}$ was taken to be
1~GeV.  The sensitivity of the results to the $k_T$ integration range was tested
by multiplying the maximum of the respective $k_T$ ranges by 0.5 and 2
respectively.

In previous estimates of intrinsic bottom production where the mass
distributions were not calculated,
average values for the transverse masses of the constituent quarks,
$m_{T \, i}^2 = m_i^2 + k_{T \, i}^2$, 
$m_{T \, q} = 0.45$ GeV and $m_{T \, b} = 4.6$ GeV were chosen
\cite{RV_SJB_ib}.  The same procedure can be employed here for
the $\Upsilon$ $x_F$ distribution which is independent of the exact value of
$m_b$ chosen.  Thus the $x_F$ distribution can be calculated
assuming simple coalescence of the $b$ and $\overline b$ in a single state,
represented in 
Eq.~(\ref{icdenom5}) by the addition of a delta function,
$\delta(x_F - x_4 - x_{5})$, in the longitudinal direction, ignoring the $k_T$
integrations where 4 and 5 represent the $b$ and $\overline b$ quarks.
When the transverse directions are also included, employing
$\delta(k_{x\, \Upsilon} - k_{x \, 4} - k_{x \, 5})$ and 
$\delta(k_{y\, \Upsilon} - k_{y \, 4} - k_{y \, 5})$, the $x_F$ distribution is found
to be independent of the $k_T$ integration range.  

\begin{figure}
  \begin{center}
    \includegraphics[width=0.495\textwidth]{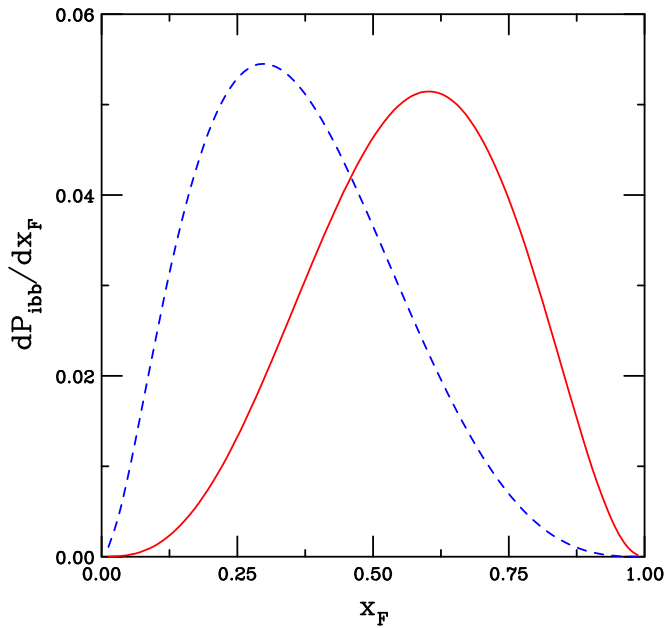}
  \end{center}
  \caption[]{(Color online) The probability distribution as a function of
    $x$ for a single $b$
    quark (dashed blue curve) and as a function of $x_F$ for
    a single $\Upsilon$ (solid red curve)
    from a five-particle Fock state of the proton.
    Both curves are normalized to unity.
  }
\label{ic_5p_xdists}
\end{figure}

Figure~\ref{ic_5p_xdists} shows the $x$ distribution of a single $b$ quark
(dashed curve) and the $x_F$ distribution of
a single $\Upsilon$ (solid curve) from
a five-particle proton Fock state.  Such a calculation provides a
check against previous results on the $J/\psi$.  The average $x$ of the $b$
quark is 0.36 from a five-particle Fock state while the average $x$ of a $c$
quark is 0.34.  Because the $c$ and $b$ quarks are both much more massive than
the light quarks in the state, there is very little difference in the average
longitudinal momentum carried by the heavy quarks.  

The $\Upsilon$ $x_F$
distribution has an average $x_F$ of 0.57, somewhat less than twice the average
$x$ of the $b$ quark.  This average is 7.5\% larger than the average $x_F$ of
the $J/\psi$ from a similar five-particle proton Fock state.
The $x_F$ 
dependence is effectively independent of the chosen $k_T$ limits, as shown in
for single $J/\psi$ production from a $|uud c\overline c \rangle$ state in
Ref.~\cite{RV_SeaQuest}.  

Following the intrinsic charm cross section calculation in
Ref.~\cite{RV_SeaQuest}, the intrinsic bottom cross section from a
$|uudb \overline b \rangle$ component of the proton can be written as 
\be
\sigma_{\rm ib \, 5}(pp) = P_{{\rm ib}\, 5}^0 \sigma_{p N}^{\rm in}
\frac{\mu^2}{4 m_{T\, b}^2} \, \, .
\label{icsign}
\ee
The factor of $\mu^2/4 m_{T \, b}^2$ arises from the soft
interaction which breaks the coherence of the Fock state.  The scale
$\mu^2 = 0.1$~GeV$^2$ is assumed for better agreement with the $J/\psi$
$A$ dependence, see Ref.~\cite{VBH1}.  An inelastic $p+N$
cross section of $\sigma_{pN}^{\rm in} = 42$~mb is 
appropriate for RHIC energies.

The $\Upsilon$ cross section can be obtained from Eq.~(\ref{icsign}) analogously
to how the cross section is obtained in perturbative QCD using the color
evaporation model (CEM) which relates the $Q \overline Q$ cross section to the
quarkonium cross sections.  The same factor, $F_B = 0.022$, determined from
recent CEM calculations \cite{RV_PRC2015} for inclusive prompt $\Upsilon$
production, is used here to relate the intrinsic bottom cross section to the
$\Upsilon$ cross section from the same state,
\be
\sigma_{{\rm ib} \, 5}^{\Upsilon}(pp) = F_B \sigma_{{\rm ib} \, 5}(pp) \, \, .
\label{icsigJpsi}
\ee

The nuclear dependence of the intrinsic bottom contribution is assumed to be the
same as that extracted for the nuclear surface-like component of $J/\psi$
dependence by the NA3 Collaboration \cite{NA3},
\be 
\sigma_{{\rm ib} \, 5}^{\Upsilon}(pA) =
\sigma_{{\rm ib} \, 5}^{\Upsilon}(pp) \, A^\beta \, \, 
\label{icsigJpsi_pA}
\ee
with $\beta = 0.71$ \cite{NA3} for a proton beam.  In that experiment, two
contributions to the cross section were separated, a volume-type component,
with a close-to-linear nuclear dependence, and the diffractive component,
associated with intrinsic charm in Ref.~\cite{VBH1}, with the $A$ dependence
given in Eq.~(\ref{icsigJpsi_pA}) above.  The $A$ dependence here is similar
to what one would expect if the incident proton interacted with the nuclear
surface acting as a black disc,
$A^{2/3}$ \cite{NA3,VBH1}.  Note that the nuclear
modification only affects the total rate, not the kinematic distributions.

Recall that the nuclear mass dependence in Eq.~(\ref{icsigJpsi_pA}) was determined from fixed-target proton-nucleus interactions.  The ANDY data were taken in Cu+Au collisions where any nucleon in either beam could fluctuate into an intrinsic bottom Fock state and be brought on shell by a soft gluon from the opposite beam.  

When generalizing Eq.~(\ref{icsigJpsi_pA}) to nucleus-nucleus collisions, the
nuclear dependence becomes more complicated.  However, there is no reason to
expect the $\Upsilon$ pair cross sections to be the same whether $A_t$ or $A_t$
is acting as the projectile.  Naive estimates using a Glauber model yield a
rate roughly 2.5 times larger in the Au-going direction than in the Cu-going
direction.  This suggests that performing the ANDY anaalysis separately for each
beam configuration could provide further insight into the production mechanism.
  
Thus if $A_p$ is the mass number of one nuclear beam and $A_t$ is its oppositely-directed collision partner, the nuclear dependence for an $A_p + A_t$ collision would be
\be 
\sigma_{{\rm ib} \, 5}^{\Upsilon}(A_p A_t) =
\sigma_{{\rm ib} \, 5}^{\Upsilon}(pp) \, A_p^{\beta_p} A_t^{\beta_t} \, \, 
\label{icsigJpsi_AB}
\ee
where the projectile and target nuclei could each have a different effective
exponent, $\beta_p$ and $\beta_t$ respectively.
The two contributions to the NA3 $J/\psi$ production data, the hard scattering
contribution with a volume-type, near linear $A$ dependence, and the
diffractive surface-type black disc contribution, are summed for the total
$J/\psi$ production cross section \cite{NA3}.
In $p+A$ collisions, one can define a
nuclear suppression factor by dividing the summed contributions by $A$
\cite{RV_SeaQuest}.
Similarly here a suppression factor in $A_p+A_t$ collisions can be calculated
by dividing the contributions to $\Upsilon$ production
from the hard and diffractive parts by the nuclear overlap
function, $T(A_p A_t)$, and the production cross section in $p+p$ collisions
\cite{RV_PRC2015}.  Note that the focus
here is only on intrinsic bottom production,
hard scattering $\Upsilon$ production in perturbative QCD has been calculated
previously and is centered at midrapidity.

The probability for intrinsic bottom production
from a $|uudb \overline b \rangle$ state
can be assumed to scale with the square of the quark mass relative to the
intrinsic charm probability,
\be
P_{{\rm ib}\, 5}^0 = P_{{\rm ic}\, 5}^0 \left(\frac{m_c^2}{m_b^2}\right) \, \, .
\label{IB_prob}
\ee
In Ref.~\cite{RV_SeaQuest}, the range $0.1 \% \leq P_{{\rm ic}\, 5}^0 \leq 1\%$
was studied.  Assuming this range,
$P_{{\rm ib}\, 5}^0 \approx 0.075 P_{{\rm ic}\, 5}^0$ for $m_c = 1.27$~GeV
\cite{NVF} with $m_b = 4.65$~GeV, the 1S value of the bottom quark mass
\cite{PDG}.
Employing this mass value gives
$0.008\% \leq P_{{\rm ib}\, 5}^0 \leq 0.08\%$.  Choosing a lower bottom quark mass
would increase the probability for intrinsic bottom production in the
five-particle Fock state.

Given this probability in Eq.~(\ref{icsign}), the intrinsic bottom cross section
in this state, $\sigma_{{\rm ib} \, 5}(pp)$, is in the range 3.6-36~nb for
$m_b = 4.65$~GeV.  Taking a lower value of the bottom quark mass in the
calculation of $\sigma_{{\rm ib} \, 5}(pp)$
would increase the cross section as well
since $m_b$ is a factor in both $P_{{\rm ib} \, 5}^0$ and the factor
$\mu^2/(4 m_{T\, b}^2)$ in Eq.~(\ref{icsign}). The total
$b \overline b$ production cross section at next-to-leading order in
perturbative QCD is $2100^{+400}_{-300}$~nb \cite{RV_something}.  The scale
factor for $\Upsilon$ production in next-to-leading order perturbative QCD is
assumed to be the same for $\Upsilon$ production from the five-particle
intrinsic bottom Fock state.  Thus the difference in cross sections for
$\Upsilon$ production in the two processes remains similar for $\Upsilon$
production relative to the total $b \overline b$ cross section.  It is, however,
worth noting that the $\Upsilon$ rapidity distribution from intrinsic bottom
has its maximum where the $\Upsilon$ cross section calculated in perturbative
QCD is steeply falling.  Thus, at forward rapidity, the $\Upsilon$ cross section
from intrinsic bottom could be comparable to or larger than the perturbative
QCD cross section.

\subsection{Double $\Upsilon$ Production from a $|uud b \overline b b \overline b \rangle$ State}
\label{double_Ups}

All the calculations in this section assume production of one or two $\Upsilon$s
from a seven-particle Fock state in a proton,
$|uud b \overline b b \overline b \rangle$.
Building on the distributions from a five-particle Fock state in
Eq.~(\ref{icdenom5}),
the frame-independent probability distribution of a $7$-particle
$b \overline b$ Fock state in the proton is 
\be
dP_{{\rm ibb}\, 7} = P_{{\rm ibb}\,7}^0
N_7 \int dx_1 \cdots dx_7 \int dk_{x\, 1} \cdots dk_{x \, 7}
\int dk_{y\, 1} \cdots dk_{y \, 7} 
\frac{\delta(1-\sum_{i=1}^7 x_i)\delta(\sum_{i=1}^7 k_{x \, i}) \delta(\sum_{i=1}^7 k_{y \, i})}{(m_p^2 - \sum_{i=1}^7 (m_{T \, i}^2/x_i) )^2} \, \, ,
\label{icdenom7}
\ee
where $i = 1$, 2, 3 are the interchangeable light quarks ($u$, $u$, $d$)
and $i = 4-7$ are the $b$ and $\overline b$ quarks respectively.
Here $N_7$ normalizes the
$|uud b \overline b b \overline b \rangle$ probability to unity and
$P_{{\rm ibb}\, 7}^0$
scales the unit-normalized distribution to the assumed probability for
production of this state.

The bottom quark mass, $m_b = 4.65$~GeV, used for the cross section estimates
in the previous section, is the 1S bottom mass
\cite{PDG}.  The mass threshold for the
double $\Upsilon$ or $b \overline b b \overline b$ state is $4m_b$ with no
internal $k_T$ of the quarks in the state.  Employing the 
1S bottom mass, $m_b = 4.65$~GeV, the mass threshold is $4m_b = 18.60$~GeV,
greater than the average ANDY mass.
Therefore, a lower limit on $m_b$ of 4~GeV is also used.
This value, 4.5\% below the running $\overline{\rm MS}$ bottom mass of
$4.18 \pm 0.03$~GeV \cite{PDG}, allows for a lower mass threshold potentially
consistent
with the ANDY measurement and the predicted tetraquark masses.  In this case,
$0.01\% \leq P_{{\rm ib}\, 5}^0 \leq 0.1\%$

The cross section for $\Upsilon$ pair production from an intrinsic
$|uud b \overline b b \overline b \rangle$ Fock state can determined analogously
to Eq.~(\ref{icsigJpsi}), but now for two $b\overline b$ pairs,
\be
\sigma_{{\rm ibb} \, 7}^{\Upsilon \Upsilon} (pp) = F_B^2 \sigma_{{\rm ibb} \, 7} (pp) \,
\, . \label{dblUps}
\ee
Equation~(\ref{icsign}) can similarly be generalized, 
assuming that no additional soft scale factor of
$\mu^2/4\hat{m}_b^2$ is needed for the second
$b \overline b$ pair in the state, as assumed in Ref.~\cite{dblic}, giving
$\sigma_{{\rm ibb}\, 7}(pp) = P_{{\rm ibb}\, 7}^0[\sigma^{\rm inel}_{pN} (\mu^2/4\hat{m}_b^2)]$.  Using Eq.~(\ref{icsign}), the term in brackets can be rewritten as
$\sigma_{{\rm ib}\, 5}/P_{{\rm ib} \, 5}^0$ so that, after substitution into
Eq.~(\ref{dblUps}), one has
\be
\sigma_{{\rm ibb} \, 7}^{\Upsilon \Upsilon} (pp) =
F_B^2 \ \frac{P_{{\rm ibb} \, 7}^0}{P_{{\rm ib} \, 5}^0}\  \sigma_{{\rm ib} \, 5} (pp)
= F_B \ \frac{P_{{\rm ibb}\, 7}^0}{P_{{\rm ib} \, 5}^0}\
\sigma_{{\rm ib}\, 5}^\Upsilon (pp) 
\label{sigib_UpsUps}
\ee
If, $P_{{\rm ibb}\, 7}^0$ scale as the mass squared relative to
$P_{{\rm icc}\, 7}^0$, as in Eq.~(\ref{IB_prob}), then
\be
P_{{\rm ibb}\, 7}^0 = P_{{\rm icc}\, 7}^0 \left(\frac{m_c^2}{m_b^2}\right)^2 \, \, ,
\label{IBB_prob}
\ee
this time with an additional power of the mass ratio squared for the second
$Q \overline Q$ pair in the Fock state.  In Ref.~\cite{dblic},
$P_{{\rm icc}\, 7}^0 \approx 4.4\% P_{{\rm ic}\, 5}^0$, based on the forward
double $J/\psi$ cross section measured by NA3 \cite{Badpi}.  Thus
$2.0 \times 10^{-4} P_{{\rm ic}\, 5}^0 \leq P_{{\rm ibb}\, 7}^0 \leq 4.5 \times 10^{-4} P_{{\rm ic}\, 5}^0$ with the lower limit obtained for $m_b = 4.65$~GeV and the
higher limit for $m_b = 4$~GeV.  Taking the higher value gives
$4.5 \times 10^{-5}\% \leq P_{{\rm ibb}\, 7}^0 \leq 4.5 \times 10^{-4} \%$,
resulting in a small expected rate.

Following Eq.~(\ref{icsigJpsi_AB}), one can write the $A_p + A_t$
  dependence of the double $\Upsilon$ cross section as
\be 
\sigma_{{\rm ibb} \, 7}^{\Upsilon \Upsilon}(A_p A_t) =
\sigma_{{\rm ibb} \, 7}^{\Upsilon \Upsilon}(pp) \, A_p^{\beta_p} A_t^{\beta_t} \, \, .
\label{dblUps_AB}
\ee
Because of the limitations in the result provided by the ANDY Collaboration,
which does not report an absolute rate, further quantiative comparisons cannot
be performed.  The remainder of
this section is thus devoted to calculations of the $\Upsilon$ pair
$x_F$ and rapidity distributions in Sec.~\ref{dblib_xy} and the pair mass
distributions in Sec.~\ref{dblib_mass}.  The goal here is to determine whether
such a $|uudb \overline b b \overline b \rangle$ state, producing either an
$\Upsilon$ pair or an $X_b(b \overline b b \overline b)$ would be compatible
with the ANDY observation, without regard to the total rate.

\subsubsection{$x_F$ and Rapidity Distributions of $\Upsilon$ Pair Production}
\label{dblib_xy}

The $x_F$ distribution for a single $\Upsilon$ from a seven-particle state
can be calculated by introducing delta functions required to coalesce one of
the $b \overline b$ pairs in the state into an $\Upsilon$,
$\delta (x_{\Upsilon \, 1} - x_{b \, 1} - x_{\overline b \, 1})$, to the probability
distribution given in Eq.~(\ref{icdenom7}).  Here the subscript `1'
simply denotes one of the two $b \overline b$ pairs in the state.  It is worth
noting that there is nothing to prevent the $\Upsilon$ from being produced by
the $b$ quark from pair 1 and the $\overline b$ quark from pair 2 (and equally
likely for the $b$ quark from pair 2 to coalesce with the $\overline b$ quark
from pair 1) since both pairs are comoving in the state.  Similar delta
functions can be applied to the transverse directions.
Such cross coalescence can add a Combinatorial factor to the probability but
will not affect the shape of the distributions.

\begin{figure}
  \begin{center}
    \includegraphics[width=0.495\textwidth]{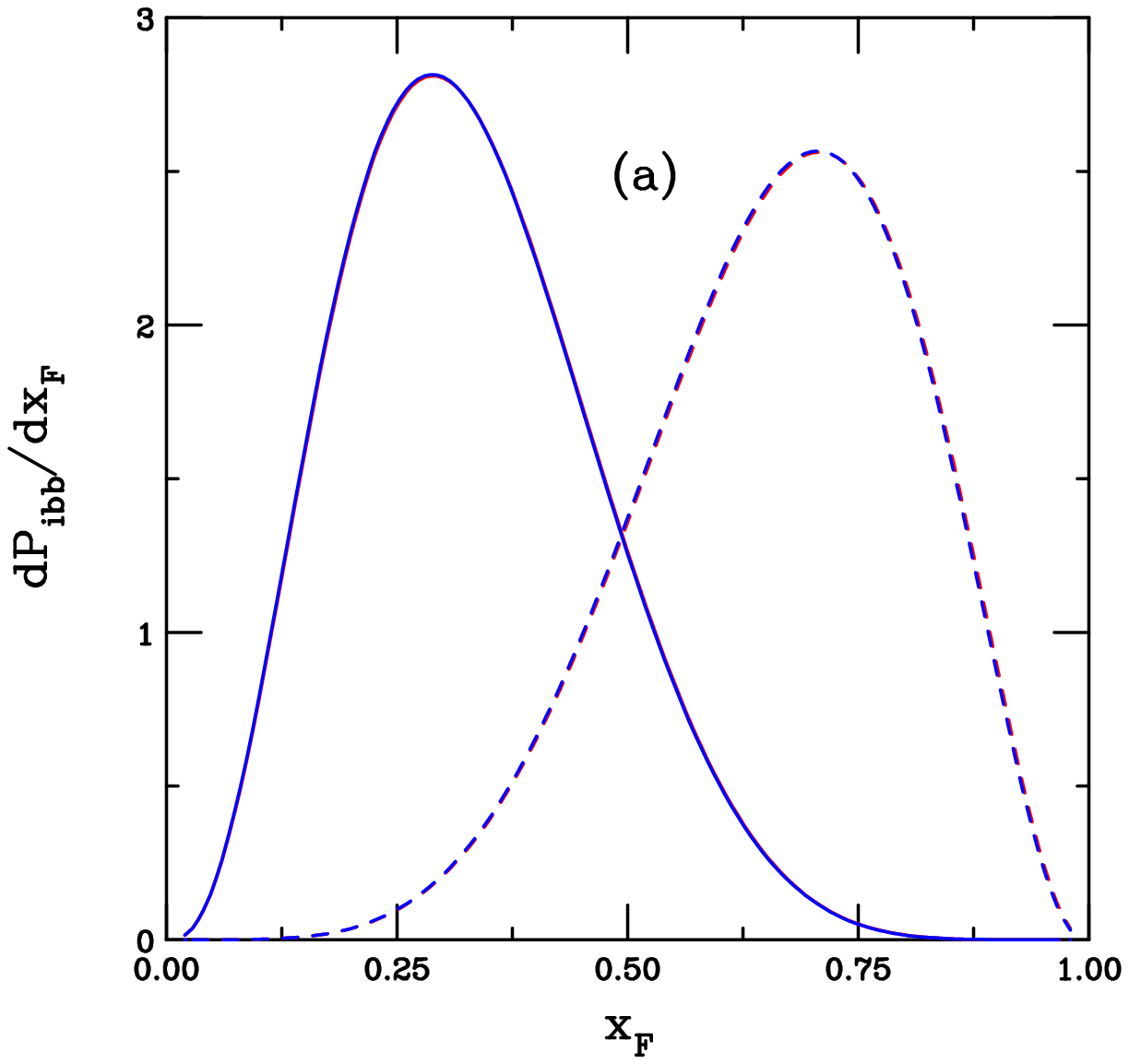}
    \includegraphics[width=0.495\textwidth]{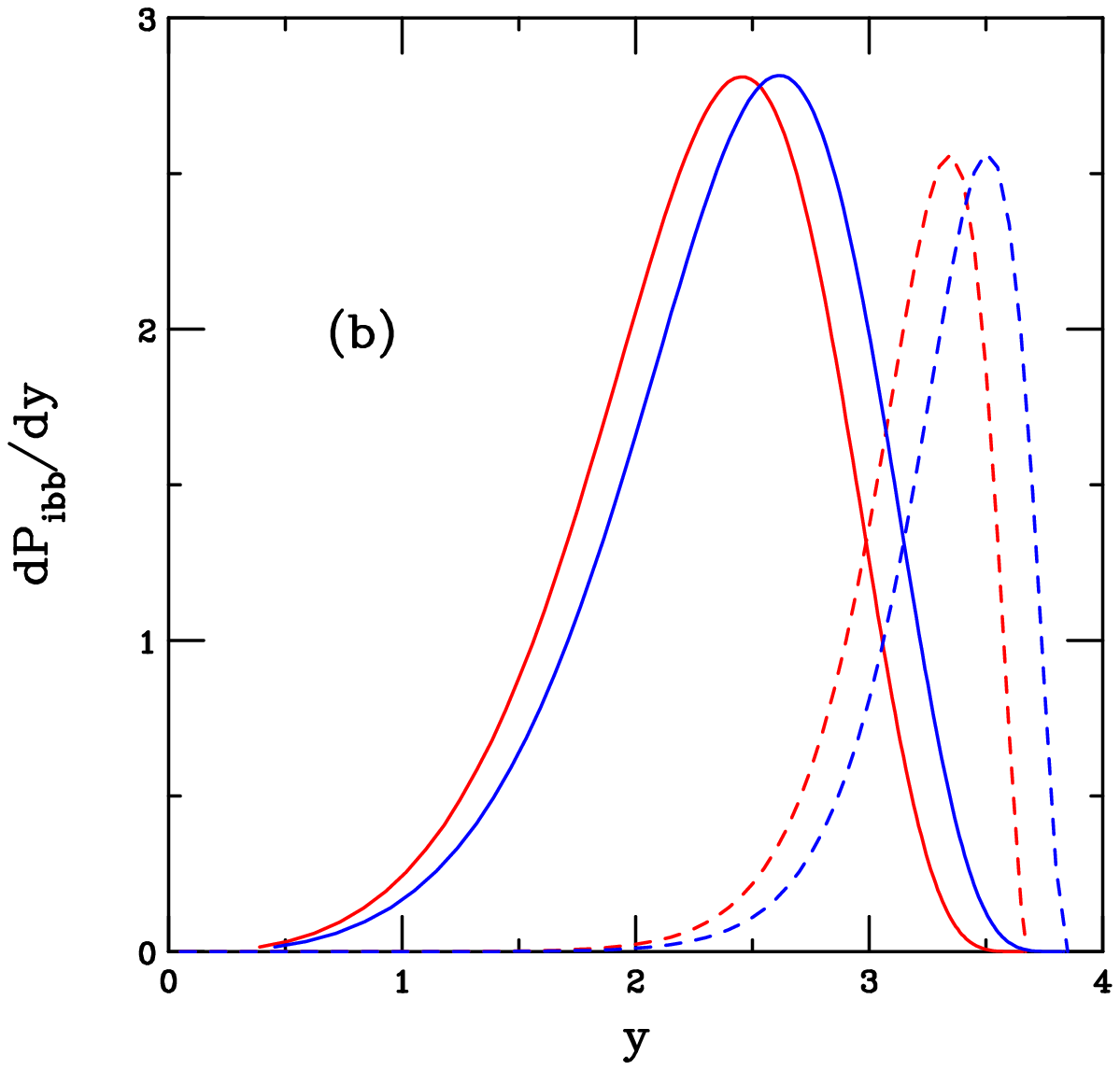}
  \end{center}
  \caption[]{(Color online) The probability distributions for a single
    $\Upsilon$ state and $\Upsilon$ pairs from a seven-particle proton Fock state
    as a function of $x_F$
    (a) and rapidity (b).  The solid curves show the single $\Upsilon$
    production from the state while the dashed curves illustrate the $\Upsilon$
    pair distributions.  The red curves employ $m_b = 4.65$~GeV while the blue
    curves display the results for $m_b = 4$~GeV.
    The results are independent of the $k_T$ integration
    range for the light and bottom quarks.  The $x_F$ distributions in (a)
    are independent of the bottom quark mass while the rapidity distributions
    exhibit an $m_b$ dependence, as illustrated in (b).  All curves are
    normalized to unity.
  }
\label{ic_xydists}
\end{figure}

Employing a seven-particle $|uud b \overline b b \overline b \rangle$ Fock
state to allow for double $\Upsilon$ production, one can expect the average
$x_F$ of a single $\Upsilon$ to be considerably reduced since the bulk of the
momentum has to now be distributed among four bottom quarks.  As shown in
Fig.~\ref{ic_xydists}(a), the single
$\Upsilon$ $x_F$ distribution from this state has an average $x_F$ of 0.33,
somewhat less than the average momentum fraction of a single $b$ quark from a
five-particle Fock state, 0.36, as shown in Fig.~\ref{ic_5p_xdists}. 

The $x_F$ distribution for a pair of $\Upsilon$s from the same Fock state
can be obtained by adding three longitudinal delta functions,
$\delta(x_{\Upsilon \Upsilon} - x_{\Upsilon \, 1} - x_{\Upsilon \, 2})\delta(x_{\Upsilon \, 1} - x_{4} - x_{5})\delta(x_{\Upsilon \, 2} - x_{6} - x_{7})$,
as well as those in
the corresponding transverse directions to Eq.~(\ref{icdenom7}).
Note again that the $b$ and $\overline b$ that coalesce into an $\Upsilon$
do not have to come from the same pair since both pairs are comoving at
similar velocities in the state.  The $\Upsilon$ pair $x_F$ distribution is
shown in the dashed curve of Fig.~\ref{ic_xydists}(a).
The average $x_F$ of the $\Upsilon$ pair is double that of a single $\Upsilon$
state, 0.66.

Both of these distributions, the single $\Upsilon$ and the $\Upsilon$ pair, are
independent of the $k_T$ range of integration as well as the $b$ quark mass
employed.  Curves with different limits of $k_T$ integration and values of $m_B$
are superimposed in the figure.

Since the ANDY Collaboration observed their signal at forward rapidity,
Fig.~\ref{ic_xydists}(b) shows the same distributions
as in Fig.~\ref{ic_xydists}(a) but now as a function of rapidity.  Although
the results are still independent of the $k_T$ integration range,
now there is a separation between peaks for different bottom quark masses with
the distribution for the lighter bottom quark mass, 4~GeV, peaking at higher
rapidity than for $m_b = 4.65$~GeV by $\approx 0.15$ units of rapidity,
$\langle y_\Upsilon \rangle = 2.26$ and 2.41, respectively.  It can
also be seen that the relatively large difference between the single and double
$\Upsilon$ distributions as a function of $x_F$ is reduced
when the distributions are viewed as a function of rapidity.  The average
$\Upsilon$ pair rapidity distribution is $\approx 0.9$ units of rapidity
greater than that
for a single $\Upsilon$ when the single $\Upsilon$ arises from a seven-particle
Fock state.  The averages are $\langle y_{\Upsilon \Upsilon} \rangle = 3.17$
and 3.33 respectively.
Thus the peak of the double $\Upsilon$ rapidity distribution is within the
range of the ANDY measurement.

\subsubsection{$\Upsilon$ Pair Mass Distributions}
\label{dblib_mass}

The $\Upsilon$ pair
mass distribution predicted from a
$|uud b \overline b b \overline b \rangle$ Fock state is
\be
\lefteqn{\frac{dP_{{\rm ibb}\, 7}}{dM^2_{\Upsilon \Upsilon}} =
  \int \frac{dx_{\Upsilon_j}}{x_{\Upsilon_j}} \frac{dx_{\Upsilon_2}}{x_{\Upsilon_2}}
  \int dm_{\Upsilon_1}^2 dm_{\Upsilon_2}^2
  \int dk_{x \, \Upsilon_1} dk_{y \, \Upsilon_1}dk_{x \, \Upsilon_2} dk_{y \, \Upsilon_2}
  \int \frac{dx_{\Upsilon \Upsilon}}{x_{\Upsilon \Upsilon}}\int dk_{x \, \Upsilon \Upsilon}
  dk_{y \, \Upsilon \Upsilon}\ dP_{{\rm ibb}\, 7}} \label{2Ups_mass} \\ 
  &   & \mbox{} \times \delta \left( \frac{m^2_{T, \Upsilon_1}}{x_{\Upsilon_1}} -
  \frac{m_{T \, 4}^2}{x_{4}} - \frac{m_{T \, 5}^2}{x_{5}} \right)
  \delta(k_{x \, 4} + k_{x \, 5} - k_{x \, \Upsilon_1})
  \delta(k_{y \, 4} + k_{y \, 5} - k_{y \, \Upsilon_1}) 
  \delta(x_{\Upsilon_1} - x_4 - x_{5}) \nonumber \\
  &   & \mbox{} \times \delta \left( \frac{m^2_{T, \Upsilon_2}}{x_{\Upsilon_2}} -
  \frac{m_{T \, 6}^2}{x_{6}} - \frac{m_{T \, 7}^2}{x_{7}} \right)
  \delta(k_{x \, 6} + k_{x \, 7} - k_{x \, \Upsilon_2})
  \delta(k_{y \, 6} + k_{y \, 7} - k_{y \, \Upsilon_2})
  \delta(x_{\Upsilon_2} - x_6 - x_{7}) \nonumber \\
&  & \mbox{} \times\delta
\left( \frac{M^2_{T, \Upsilon \Upsilon}}{x_{\Upsilon \Upsilon}} - \frac{m_{T,
\Upsilon_1}^2}{x_{\Upsilon_1}} - \frac{m_{T, \Upsilon_2}^2}{x_{\Upsilon_2}} \right)
\delta(k_{x \, \Upsilon_1} + k_{x \, \Upsilon_2} -
k_{x \, \Upsilon  \Upsilon})
\delta(k_{y \, \Upsilon_1} + k_{y \, \Upsilon_2} -
k_{y\, \Upsilon  \Upsilon})
\delta(x_{\Upsilon \Upsilon} - x_{\Upsilon_1} - x_{\Upsilon_2}) \, \, , \nonumber
\ee
where $dP_{{\rm ibb}\, 7}$ is taken from Eq.~(\ref{icdenom7}).  The pair mass
distributions require integration over the invariant mass of each $\Upsilon$,
$2m_b < m_{T \, \Upsilon} < 2m_B$, including its momentum fraction, $x_\Upsilon$,
and its transverse momenta, as
well as integration over the $x$ and $k_T$ of the pair itself.
The delta functions insure conservation of momentum for both $\Upsilon$
mesons and the $\Upsilon \Upsilon$ pair.  

\begin{figure}
  \begin{center}
    \includegraphics[width=0.495\textwidth]{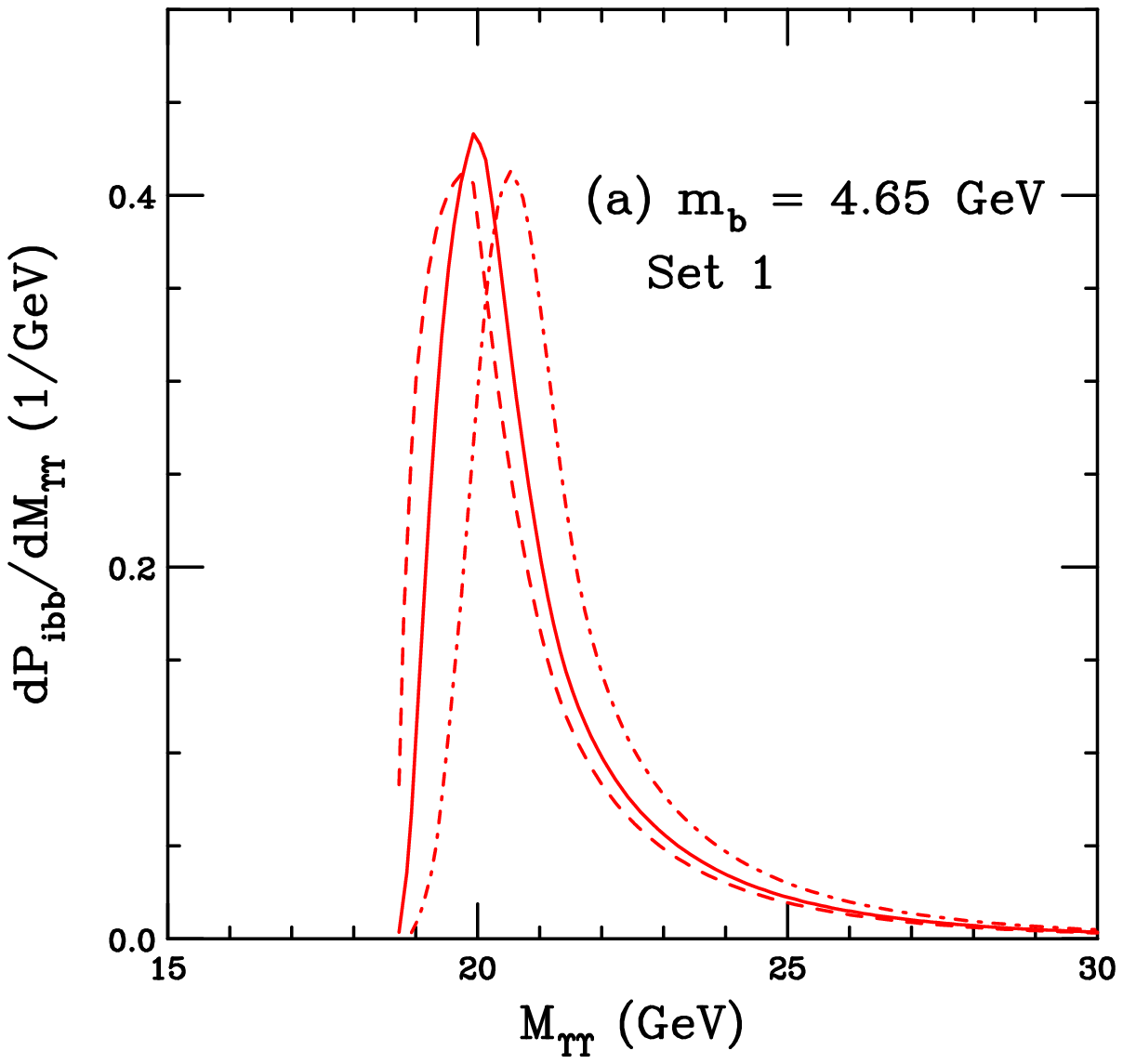}
    \includegraphics[width=0.495\textwidth]{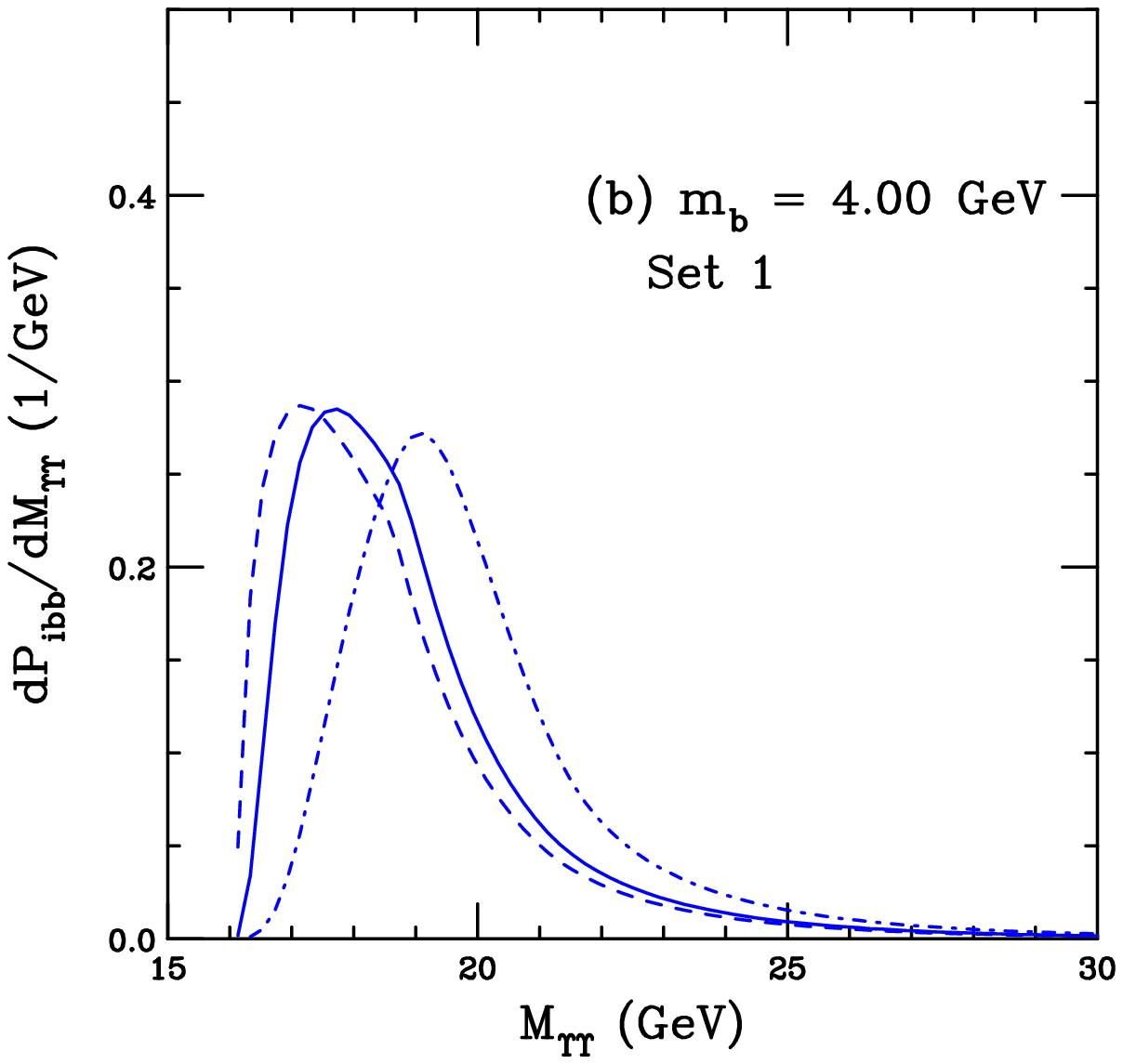}
  \end{center}
  \caption[]{(Color online) The probability for double $\Upsilon$ production
    from a seven-particle Fock state as a function of the pair mass for three
    different $k_T$ integration ranges, corresponding to Set 1 in
    Table~\protect\ref{ave_mass_table}, are shown: $k_q^{\rm max} = 0.2$~GeV,
    $k_b^{\rm max} = 1.0$~GeV and $k_\Upsilon^{\rm max} = 1.0$~GeV (solid);
    $k_q^{\rm max} = 0.1$~GeV, $k_b^{\rm max} = 0.5$~GeV and
    $k_\Upsilon^{\rm max} = 0.5$~GeV (dashed); and
    $k_q^{\rm max} = 0.4$~GeV, $k_b^{\rm max} = 2.0$~GeV and
    $k_\Upsilon^{\rm max} = 2.0$~GeV (dot-dashed).
    All distributions are normalized to unity.
    In (a) $m_b = 4.65$~GeV while in (b) $m_b = 4.0$~GeV.
  }
\label{ic_Mdists}
\end{figure}

Figure~\ref{ic_Mdists} shows the predictions for the $\Upsilon \Upsilon$
pair mass distributions.  All distributions are normalized to
unity.  Without any $k_T$ dependence,
the pair mass distribution is strongly peaked at the $4m_b$ threshold.
The $k_T$
dependence smears out the pair distribution, increasing
$\langle M_{\Upsilon \Upsilon} \rangle$ by several GeV.
The chosen default values of the transverse momentum range,
$k_q^{\rm max} = 0.2$~GeV and $k_b^{\rm max} = 1.0$~GeV, as also assumed for light
quarks and charm quarks respectively in a five-particle proton Fock state for
$J/\psi$ production \cite{RV_SeaQuest}.  In addition, the $k_\Upsilon$
integration range also is required to calculate the $\Upsilon$ pair mass
distribution.  The value $k_\Upsilon^{\rm max} = 1$~GeV is used as a default
for both the $\Upsilon$ $k_T$ range as well as that for the $\Upsilon$ pair.
The pair mass distributions for $m_b = 4.65$~GeV, the bottom quark mass used in
bottom quark calculations at next-to-leading order \cite{RVazi2},
are shown in the solid curve
of Fig.~\ref{ic_Mdists}(a).

The mass threshold is well above the average
$\Upsilon$ pair mass of 18.15~GeV reported by the ANDY
Collaboration \cite{ANDY}.  The
average pair mass in all cases is greater than 20~GeV.  Varying
the transverse integration range by increasing (dot-dashed curve) or decreasing
(dashed curve) the range
by a factor of two does not change the mass threshold and alters
the average pair mass by less than 0.5~GeV, see the results labeled Set 1 on
the left-hand side of Table~\ref{ave_mass_table}.

The pair mass obtained with the 1S value of $m_b$
is much higher than the ANDY result.   The lower value of the
bottom quark mass, 4~GeV, used with the
same set of transverse momentum integration ranges, does result in a lower
average mass, as shown in
Fig.~\ref{ic_Mdists}(b).
Even though the threshold is reduced and the average pair mass is decreased
to $\approx 19$~GeV, see the upper results on the right-hand side of
Table~\ref{ave_mass_table}, these values are still higher than obtained by
the ANDY Collaboration.  

In Ref.~\cite{dblic}, predictions for double $\Upsilon$ production from the
seven-particle Fock state were given assuming $m_{T \, b} = 4.6$ GeV.  (In that
work, only averages were reported, no distributions were presented, and no
systematic studies of the mass and transverse momentum were carried out.)
It was found that the single $\Upsilon$ and $\Upsilon \Upsilon$ pair $x$
distributions were similar to the equivalent $J/\psi J/\psi$
distributions.  The average mass, $\langle M_{\Upsilon \Upsilon} \rangle$,
was found to be 21.7~GeV for a
proton beam, a few GeV above the two $\Upsilon$ mass threshold,
$2m_\Upsilon = 18.9$ GeV.  These results are in good agreement with the central
value obtained for $m_b = 4.65$~GeV shown in Fig.~\ref{ic_Mdists}(a) and in
Table~\ref{ave_mass_table}.

The $\Upsilon$ pair mass distributions are quite sensitive to
the range of $k_T$ integration.
Results for other values of $k_q^{\rm max}$, $k_b^{\rm max}$ and
$k_\Upsilon^{\rm max}$ are shown for both values of $m_b$ considered.
Sets 2 through 5 in Table~\ref{ave_mass_table} give
the average $\Upsilon$ pair mass for each
central value as well as for halving and doubling the ranges for each chosen
central value.  Note that the central value of $k_q^{\rm max}$ is always 0.2~GeV.
The central value of $k_b^{\rm max}$ is varied from 1 to 2~GeV and, finally,
the central value of $k_\Upsilon^{\rm max}$  is either 1, 2 or 5~GeV.  The value
of $k_b^{\rm max}$ is assumed to be either less than or equal to
$k_\Upsilon^{\rm max}$, never larger.  

The larger the upper limit of the 
range of integration over momentum, the larger the average pair mass becomes.
The bottom quark mass obtained from the $\Upsilon$(1S), $m_b = 4.65$~GeV, gives
a consistently higher mass of the double $\Upsilon$ state, always greater than
20~GeV.  On the other hand, the lower limit chosen for the $b$ quark mass,
4~GeV, results in a smaller $\Upsilon$ pair mass, as low as 18.6~GeV,
but still larger than the average $\Upsilon$ pair mass reported by ANDY.

\begin{table}
  \begin{center}
    \begin{tabular}{|c|c|c|c|c||c|c|c|c|} \hline
 Set & \multicolumn{4}{|c||}{$m_b = 4.65$~GeV} & \multicolumn{4}{|c|}{$m_b = 4.00$~GeV} \\ \hline 
   &   $k_{q}^{\rm max}$ (GeV) & $k_{b}^{\rm max}$ (GeV) & $k_{\Upsilon}^{\rm max}$ (GeV) & $\langle M_{\Upsilon \Upsilon}\rangle$ (GeV) & 
$k_{q}^{\rm max}$ (GeV) & $k_{b}^{\rm max}$ (GeV) & $k_{\Upsilon}^{\rm max}$ (GeV) & $\langle M_{\Upsilon \Upsilon}\rangle$ (GeV)  \\ \hline 
  & 0.2 & 1.0 & 1.0 & 21.02 & 0.2 & 1.0 & 1.0 & 18.96 \\
1 & 0.1 & 0.5 & 0.5 & 20.74 & 0.1 & 0.5 & 0.5 & 18.59 \\
  & 0.4 & 2.0 & 2.0 & 21.60 & 0.4 & 2.0 & 2.0 & 19.98 \\ \hline
 
  & 0.2 & 1.0 & 2.0 & 21.37 & 0.2 & 1.0 & 2.0 & 19.47 \\
2 & 0.1 & 0.5 & 1.0 & 20.87 & 0.1 & 0.5 & 1.0 & 18.76 \\
  & 0.4 & 2.0 & 4.0 & 22.12 & 0.4 & 2.0 & 4.0 & 20.82 \\ \hline
 
  & 0.2 & 2.0 & 2.0 & 21.59 & 0.2 & 2.0 & 2.0 & 19.98 \\
3 & 0.1 & 1.0 & 1.0 & 21.02 & 0.1 & 1.0 & 1.0 & 18.95 \\
  & 0.4 & 4.0 & 4.0 & 22.56 & 0.4 & 4.0 & 4.0 & 21.33 \\ \hline
 
  & 0.2 & 1.0 & 5.0 & 22.10 & 0.2 & 1.0 & 5.0 & 21.01 \\
4 & 0.1 & 0.5 & 2.5 & 21.49 & 0.1 & 0.5 & 2.5 & 19.62 \\
  & 0.4 & 2.0 & 10.0 & 22.68 & 0.4 & 2.0 & 10.0 & 22.04 \\ \hline
 
  & 0.2 & 2.0 & 5.0 & 22.32 & 0.2 & 2.0 & 5.0 & 21.17 \\
5 & 0.1 & 1.0 & 2.5 & 21.56 & 0.1 & 1.0 & 2.5 & 19.77 \\
  & 0.4 & 4.0 & 10.0 & 23.88 & 0.4 & 4.0 & 10.0 & 22.98 \\ \hline
    \end{tabular}
  \end{center}
  \caption[]{The average $\Upsilon$ pair mass for given values of the bottom quark mass and the maximum range of $k_T$ integration for light quarks, bottom quarks, and the $\Upsilon$ state.}
  \label{ave_mass_table}
\end{table}

\begin{figure}
  \begin{center}
    \includegraphics[width=0.495\textwidth]{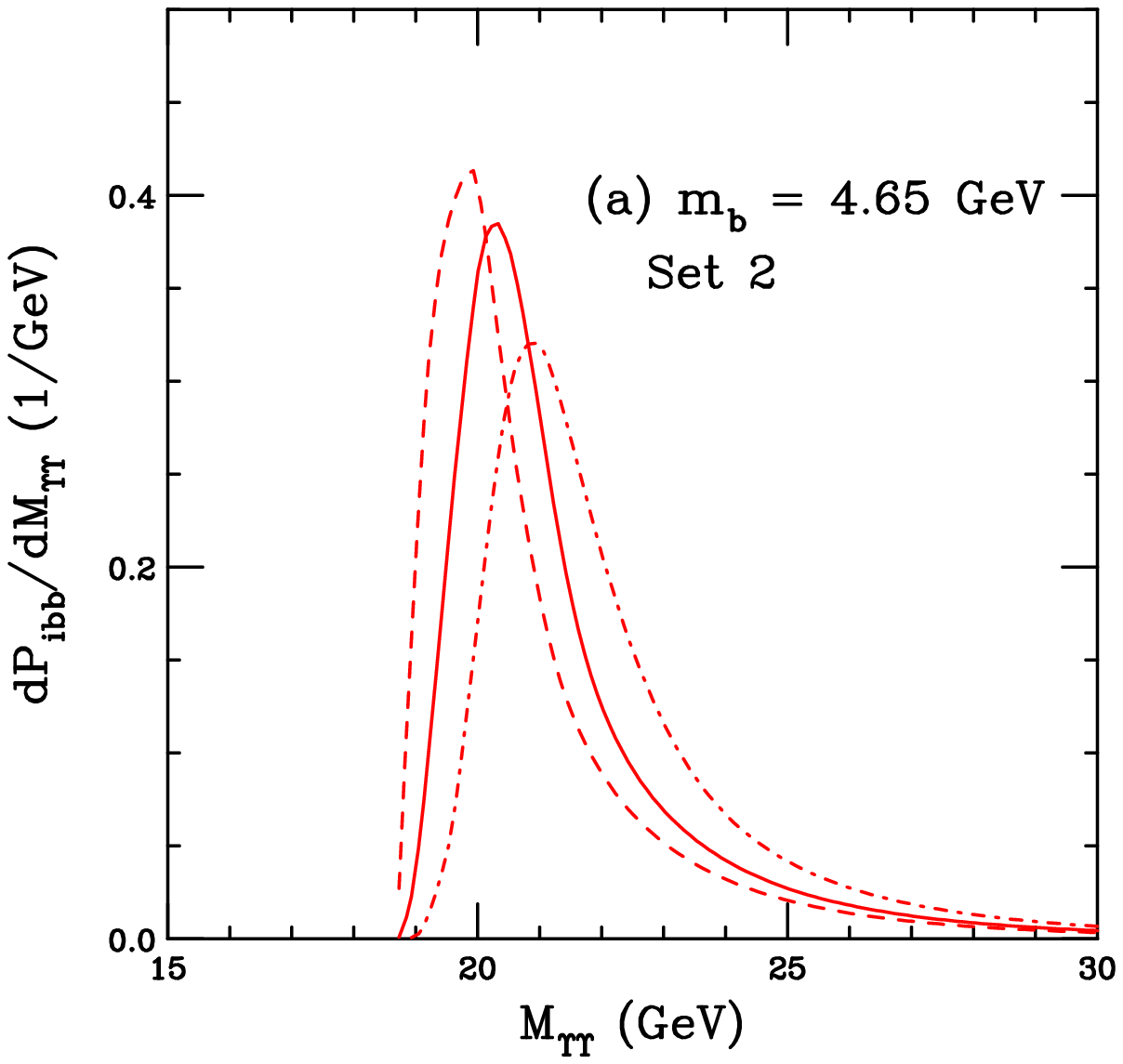}
    \includegraphics[width=0.495\textwidth]{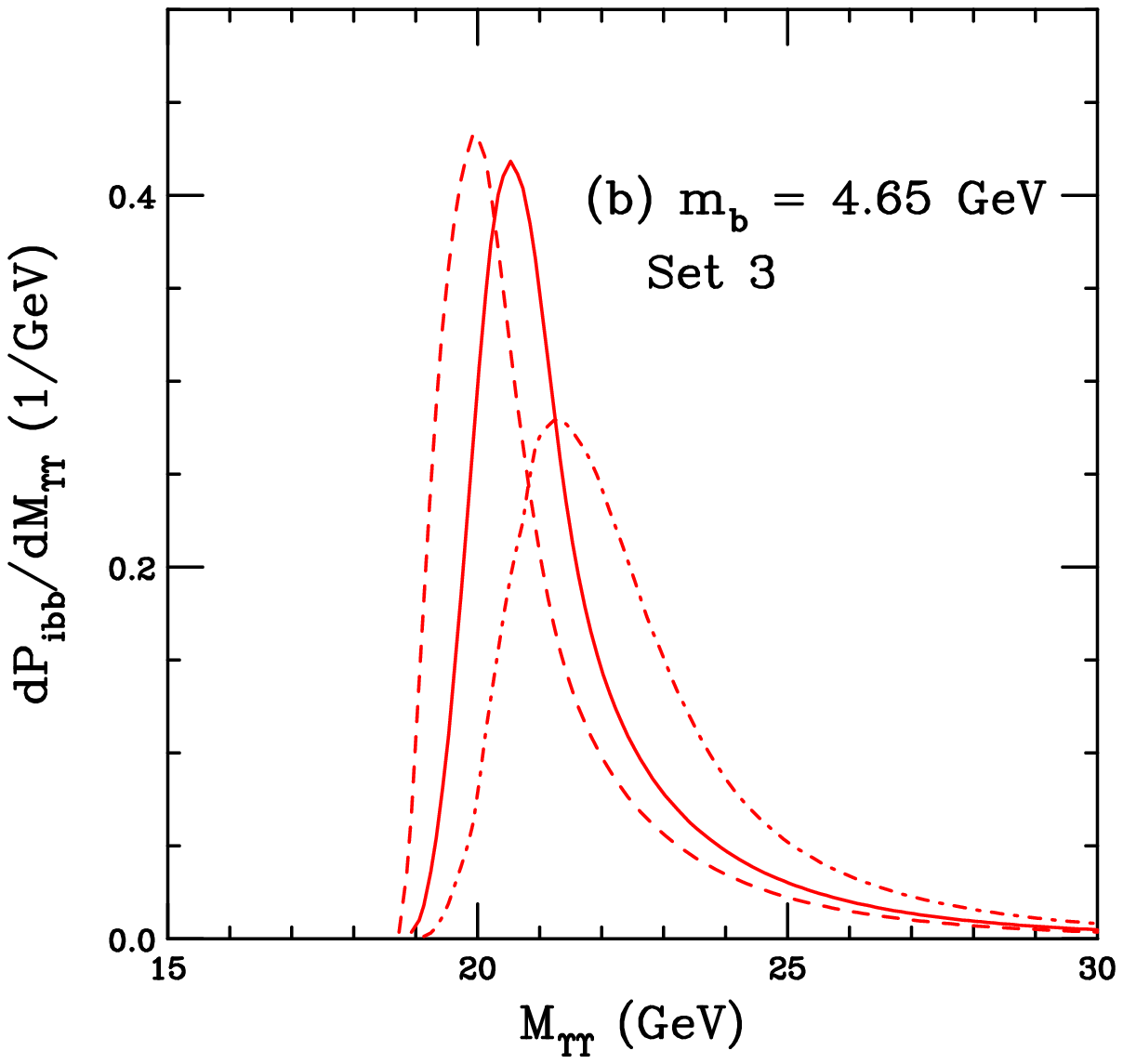}
    \includegraphics[width=0.495\textwidth]{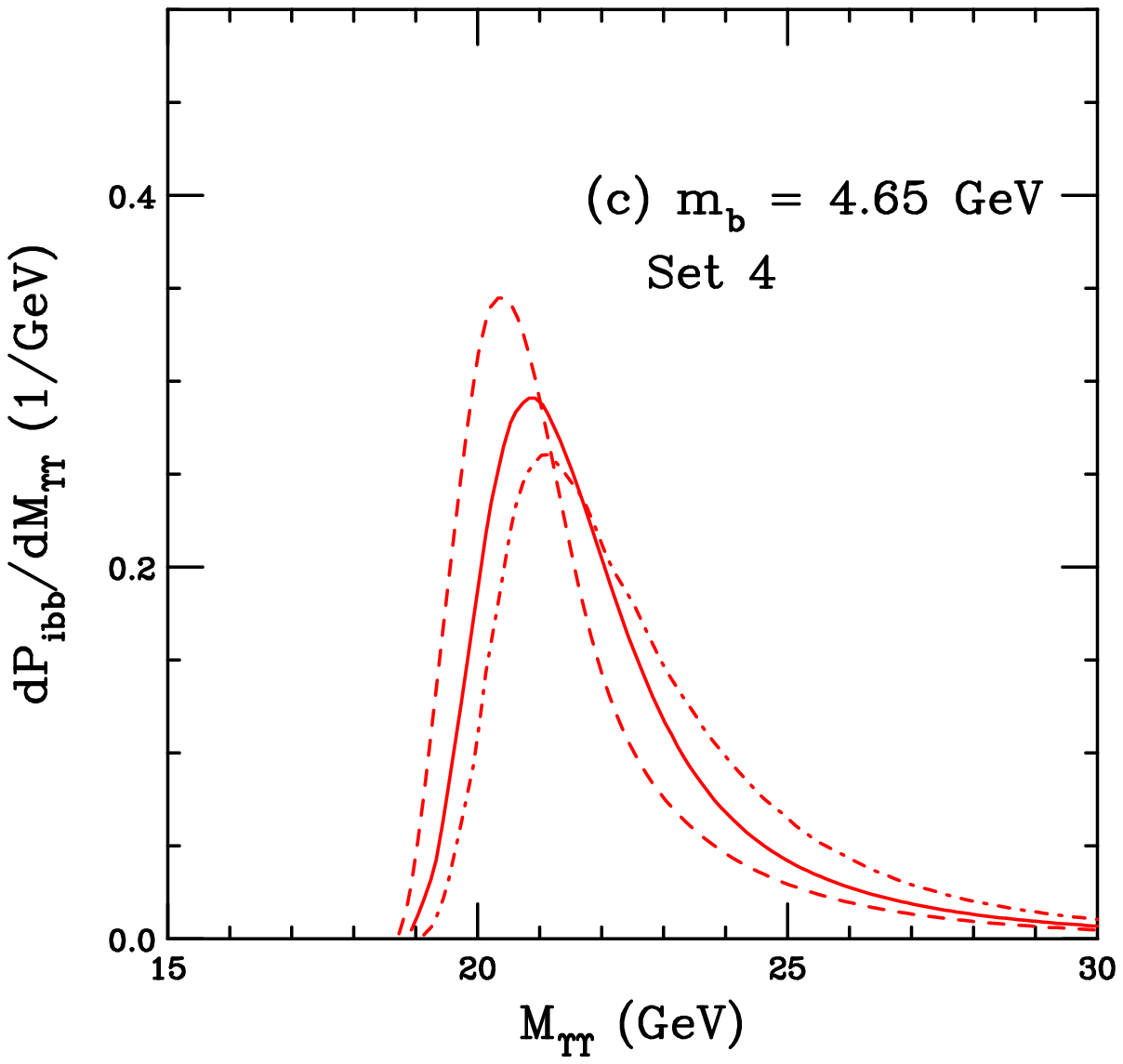}
    \includegraphics[width=0.495\textwidth]{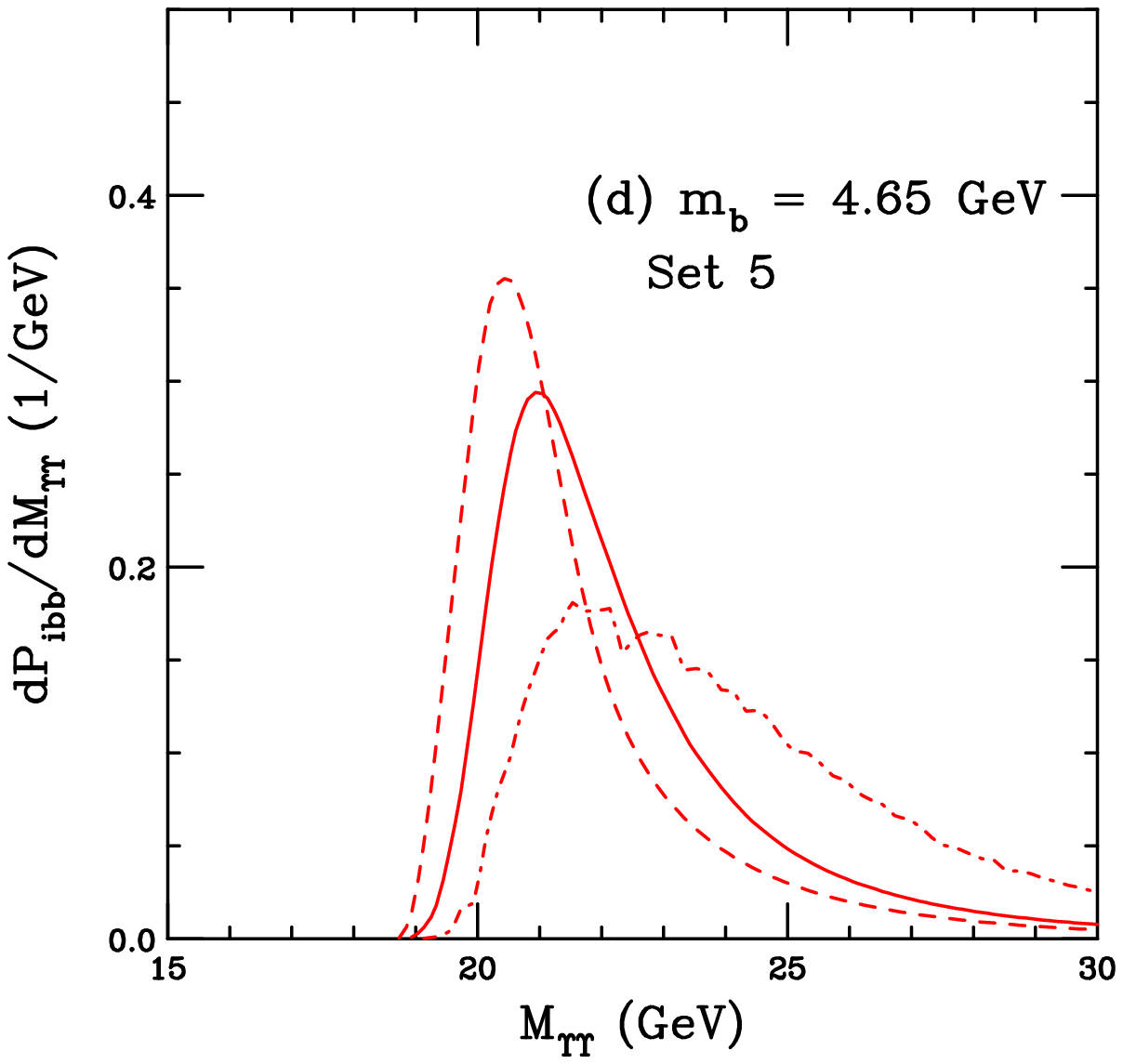}
  \end{center}
  \caption[]{(Color online) The double $\Upsilon$ pair mass distribution
    normalized to unity from a seven-particle proton Fock state with
    $m_b = 4.65$~GeV.  Different $k_T$ integration
    ranges, corresponding to Sets 2-5 of Table~\protect\ref{ave_mass_table},
    are shown in (a)-(d) respectively.
    In each case, the solid curve shows the calculation for
    the central value of the range while the dashed and dot-dashed curves
    show half and double the central value of the ranges respectively.
  }
\label{ic_Mdists_465}
\end{figure}

\begin{figure}
  \begin{center}
    \includegraphics[width=0.495\textwidth]{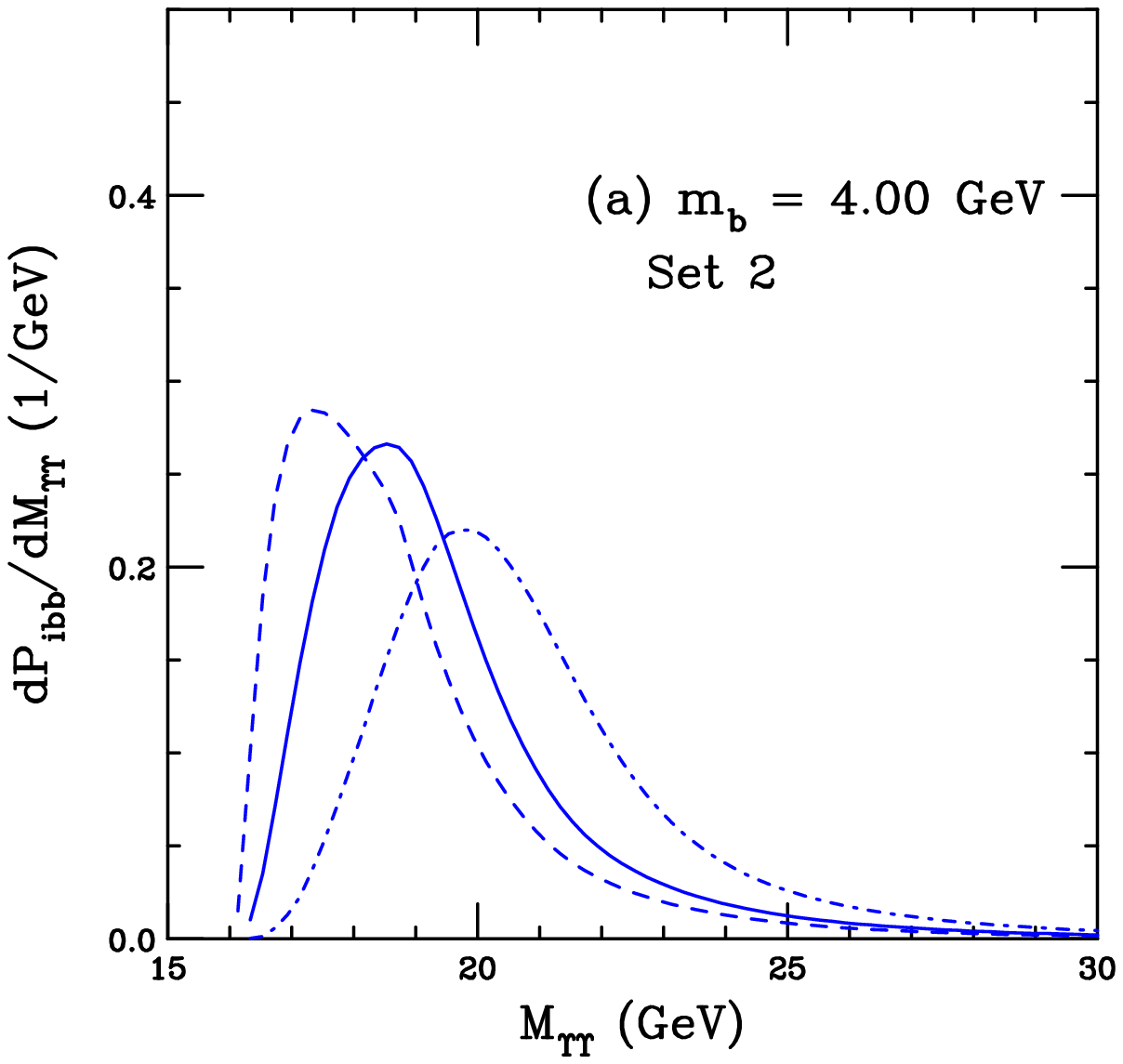}
    \includegraphics[width=0.495\textwidth]{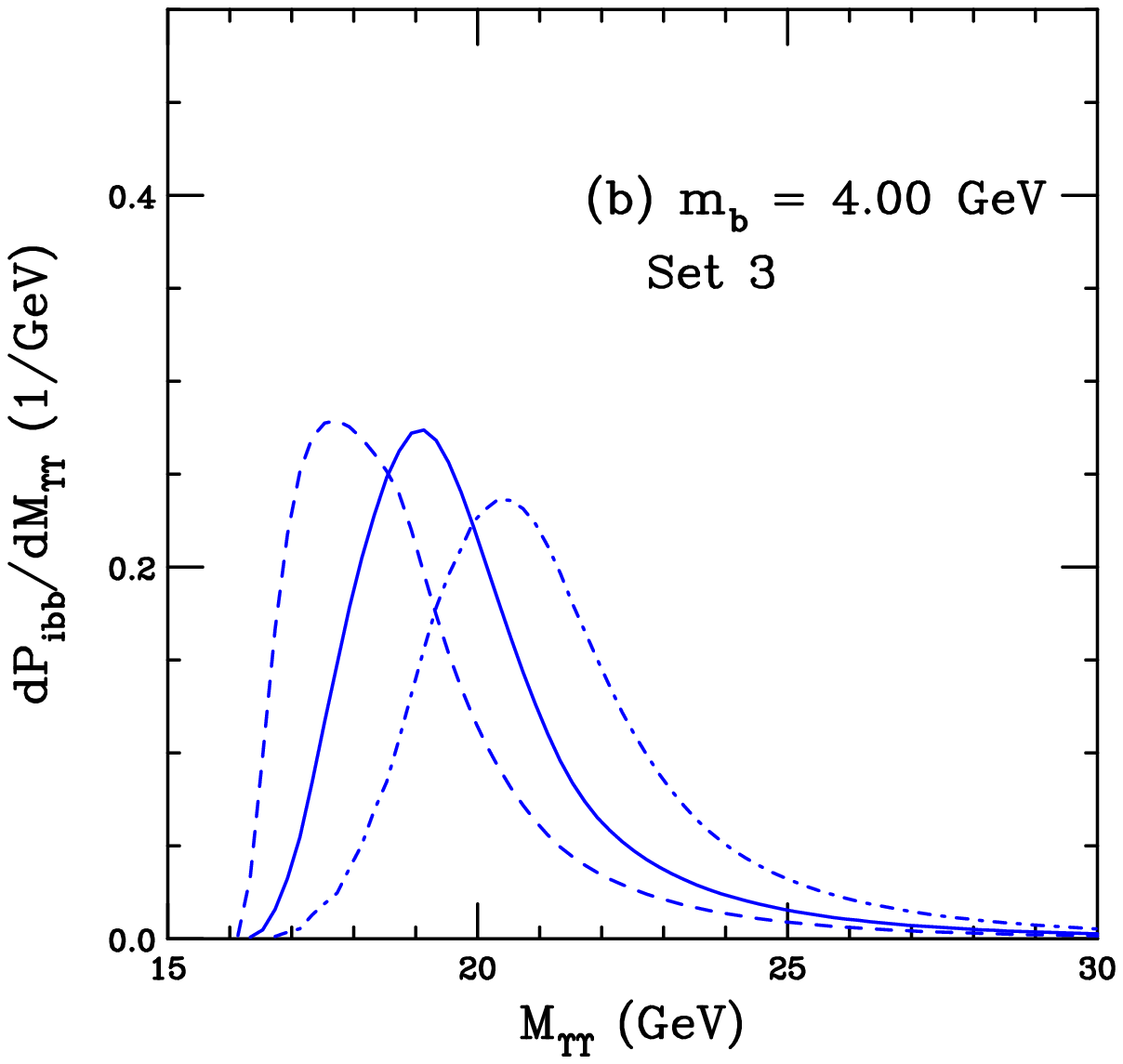}
    \includegraphics[width=0.495\textwidth]{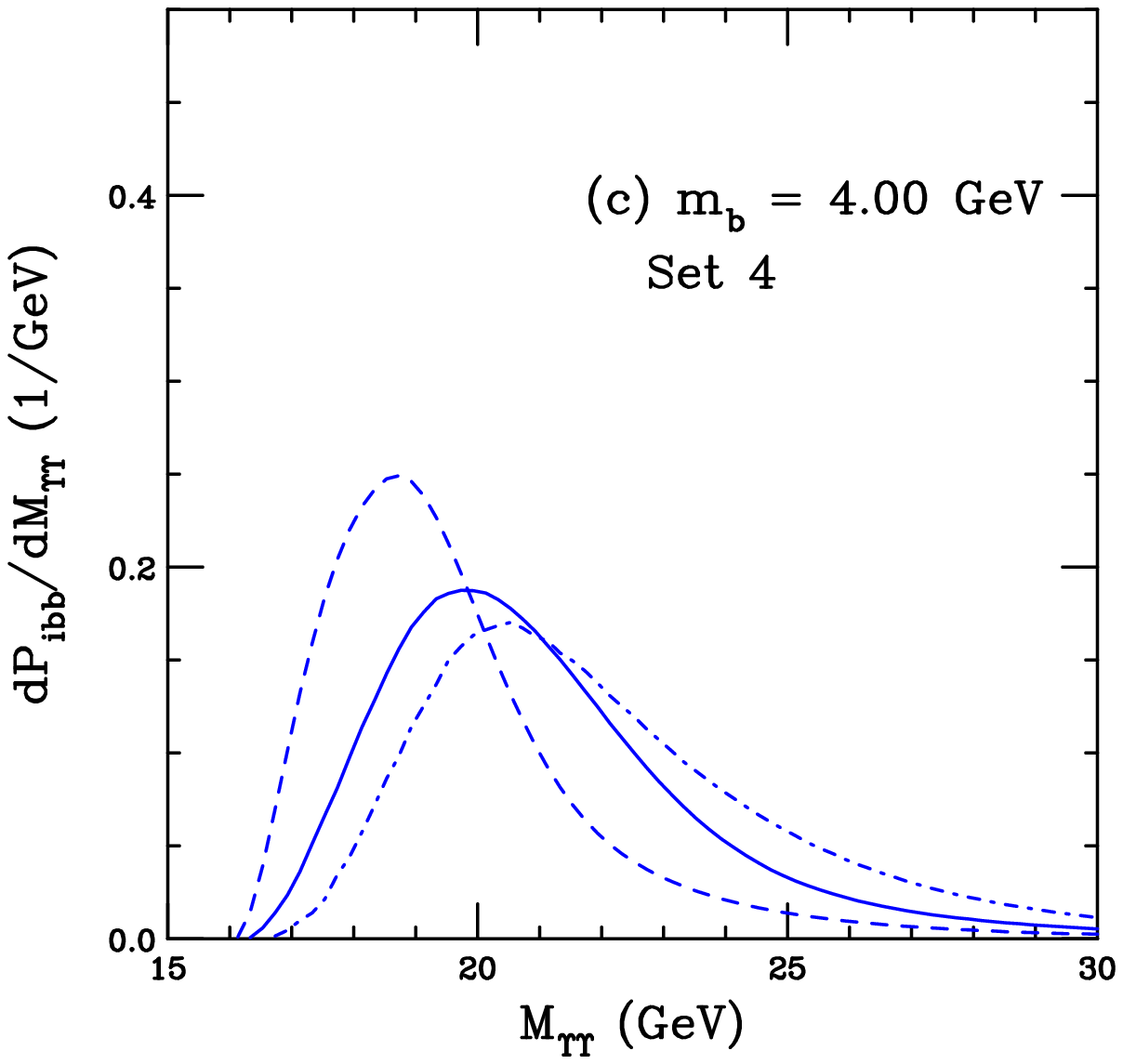}
    \includegraphics[width=0.495\textwidth]{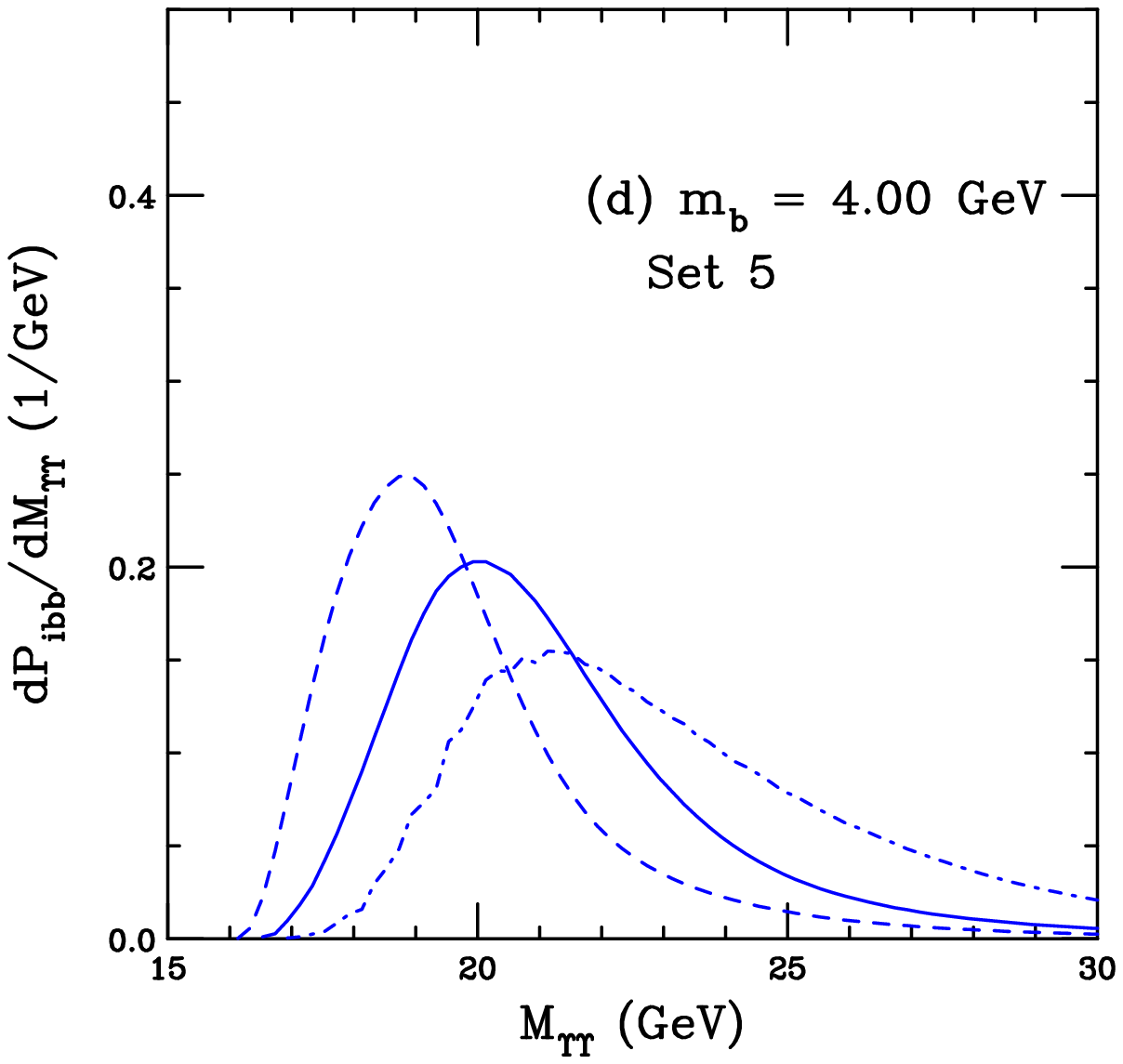}
  \end{center}
  \caption[]{(Color online) The double $\Upsilon$ pair mass distribution
    normalized to unity from a seven-particle proton Fock state with
    $m_b = 4.00$~GeV.  Different $k_T$ integration
    ranges, corresponding to Sets 2-5 of Table~\protect\ref{ave_mass_table},
    are shown in (a)-(d) respectively.
    In each case, the solid curve shows the calculation for
    the central value of the range while the dashed and dot-dashed curves
    show half and double the central value of the ranges respectively.
  }
\label{ic_Mdists_400}
\end{figure}

The pair mass distributions are shown in Figs.~\ref{ic_Mdists_465} and
\ref{ic_Mdists_400} from Sets 2-5 of $k_T$ integration ranges in
Table~\ref{ave_mass_table}.
The central values, the first line of each set in the table, are shown by the
solid curves while the dashed curve shows the result for the integration range
reduced by half and the dot-dashed curve presents the results for doubling
the $k_T$ integration range.  All distributions are normalized to unity.
It is clear that not just the average pair mass increases with the
integration range in $k_T$ but also the width of the distribution.  
In cases where half the integration range is taken (dashed curves), the width
is narrower while, when the integration range is doubled (dot-dashed curves),
the distributions are broader.

None of the results presented here are compatible with the ANDY results.
Indeed, increasing the integration range does not affect the threshold and
effectively only serves to increase the pair mass.  Unlike the earlier NA3
results on double $J/\psi$ production, where the pair mass was well above twice
the $J/\psi$ mass, the ANDY Collaboration reports an average pair
mass {\em below} twice the mass of a single $\Upsilon$(1S) state, 18.92~GeV.
Assuming $m_b = 4$~GeV in the $b \overline b$ pair mass integration,
it is possible to obtain a pair mass below $2m_\Upsilon$ but
this seems unlikely, especially since it is obtained with a particularly low
bottom quark mass, less than what is generally assumed for the current bottom
quark mass.  Indeed, in calculations of uncertainty bands on perturbative
bottom quark production such as in FONLL \cite{FONLL}, the $b$ quark mass
range is $4.5 \leq m_b \leq 5$~GeV.

It is noteworthy, however, that the average calculated pair masses, for Set 1
and $m_b = 4$~GeV, as in
Fig.~\ref{ic_Mdists}(b), are relatively compatible with the models of $X_b$
masses.  The next section considers the effects on the distributions reported
here if the internal kinematics of the Fock state assumes a single tetraquark
configuration of $b \overline b b \overline b$ not directly bound in an
$\Upsilon$ pair.

\subsection{$X_b$ Tetraquark Production from a $|uud b \overline b b \overline b \rangle$ State}
\label{Xb_tet7}

The restriction of two $\Upsilon$(1S) states produced from the
$|uud b \overline b b\overline b \rangle$ Fock state is lifted in this section
to determine whether a direct $X_b$ configuration, without the assumption of
two bound $\Upsilon$(1S) states, might be more compatible with the reported
ANDY mass distribution.

The same starting point, Eq.~(\ref{icdenom7}), is
assumed.  Now, however, the $x_F$ distribution is calculated with the addition
of a single longitudinal delta function,
$\delta(x_{X_b} - x_4 - x_5 - x_6 - x_7)$.  Despite the rearrangement of the
bottom quarks in this configuration, the $M_{X_b}$
$x_F$ distribution is identical to the one for the $\Upsilon$ pair shown
in the dashed curve in Fig.~\ref{ic_xydists}(a).  This suggests that the
longitudinal distributions are independent of any correlations between the
partons in the state as long as the same number and type of partons are
included.

Next, the mass distributions are calculated, without including the
pair mass constraints, $2m_b < m_{T \, \Upsilon} < 2m_B$,
in Eq.~(\ref{2Ups_mass}).  The $X_b$ mass
distribution is then
\be
\frac{dP_{{\rm ibb}\, 7}}{dM^2_{X_b}} & = &
  \int \frac{dx_{X_b}}{x_{X_b}}\int dk_{x \, X_b}
  dk_{y \, X_b}\ dP_{{\rm ibb}\, 7} \,\,
\delta\left( \frac{M^2_{T, X_b}}{x_{X_b}} - \frac{m_{T, 4}^2}{x_{4}}
- \frac{m_{T, 5}^2}{x_{5}}- \frac{m_{T, 6}^2}{x_{6}}- \frac{m_{T, 7}^2}{x_{7}}
\right) \label{Xb_mass}
\\ 
&  & \mbox{} \times 
\delta(k_{x \, 4} + k_{x \, 5} + k_{x \, 6} + k_{x \, 7}
- k_{x \, X_b})
\delta(k_{y \, 4} + k_{y \, 5} + k_{y \, 6} + k_{y \, 7}
- k_{y \, X_b})
\delta(x_{X_b} - x_4 - x_5 - x_6 - x_7) \, \, , \nonumber
\ee
where $dP_{{\rm ibb}\, 7}$ is from Eq.~(\ref{icdenom7}). 

\begin{figure}
  \begin{center}
    \includegraphics[width=0.495\textwidth]{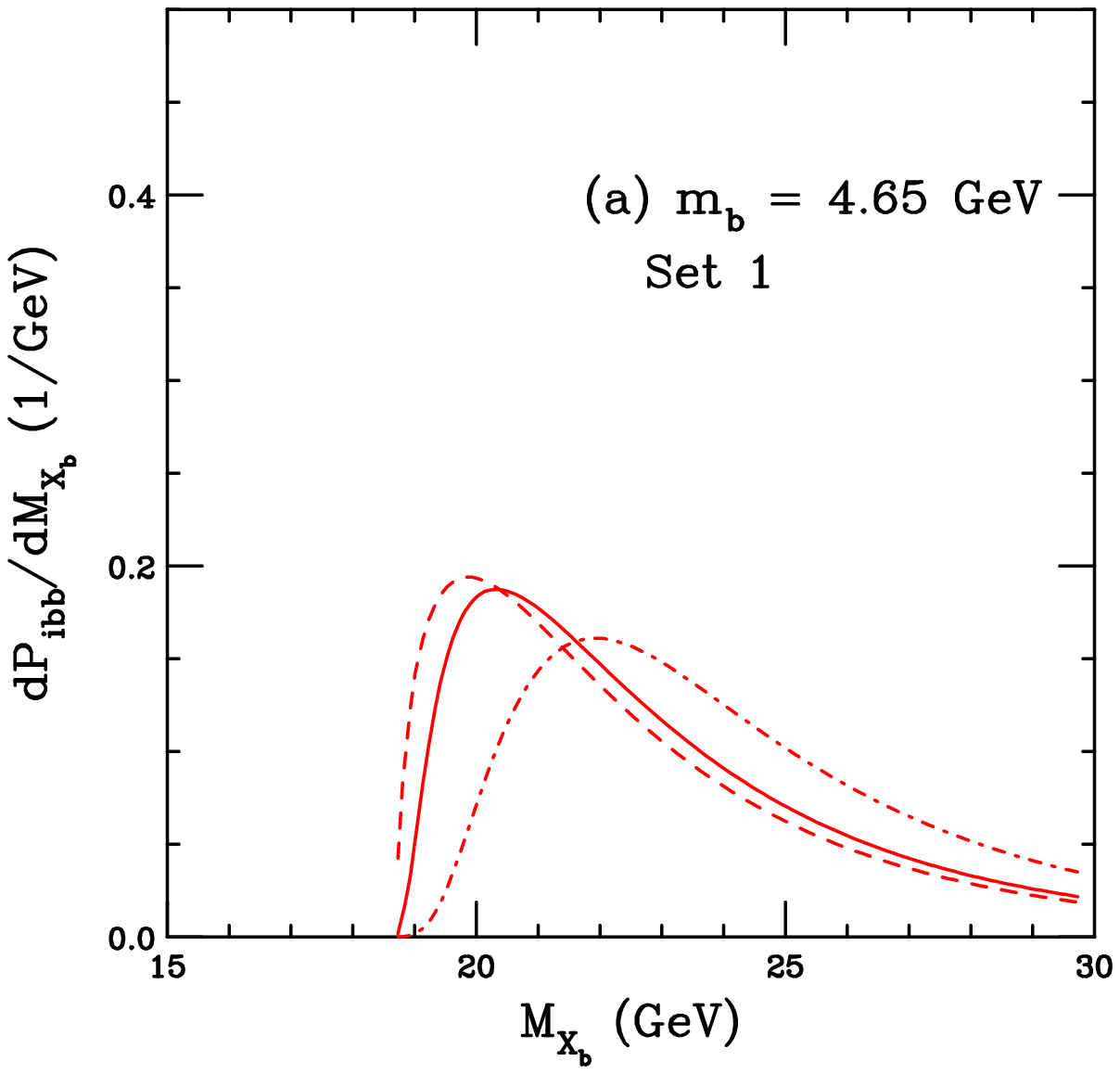}
    \includegraphics[width=0.495\textwidth]{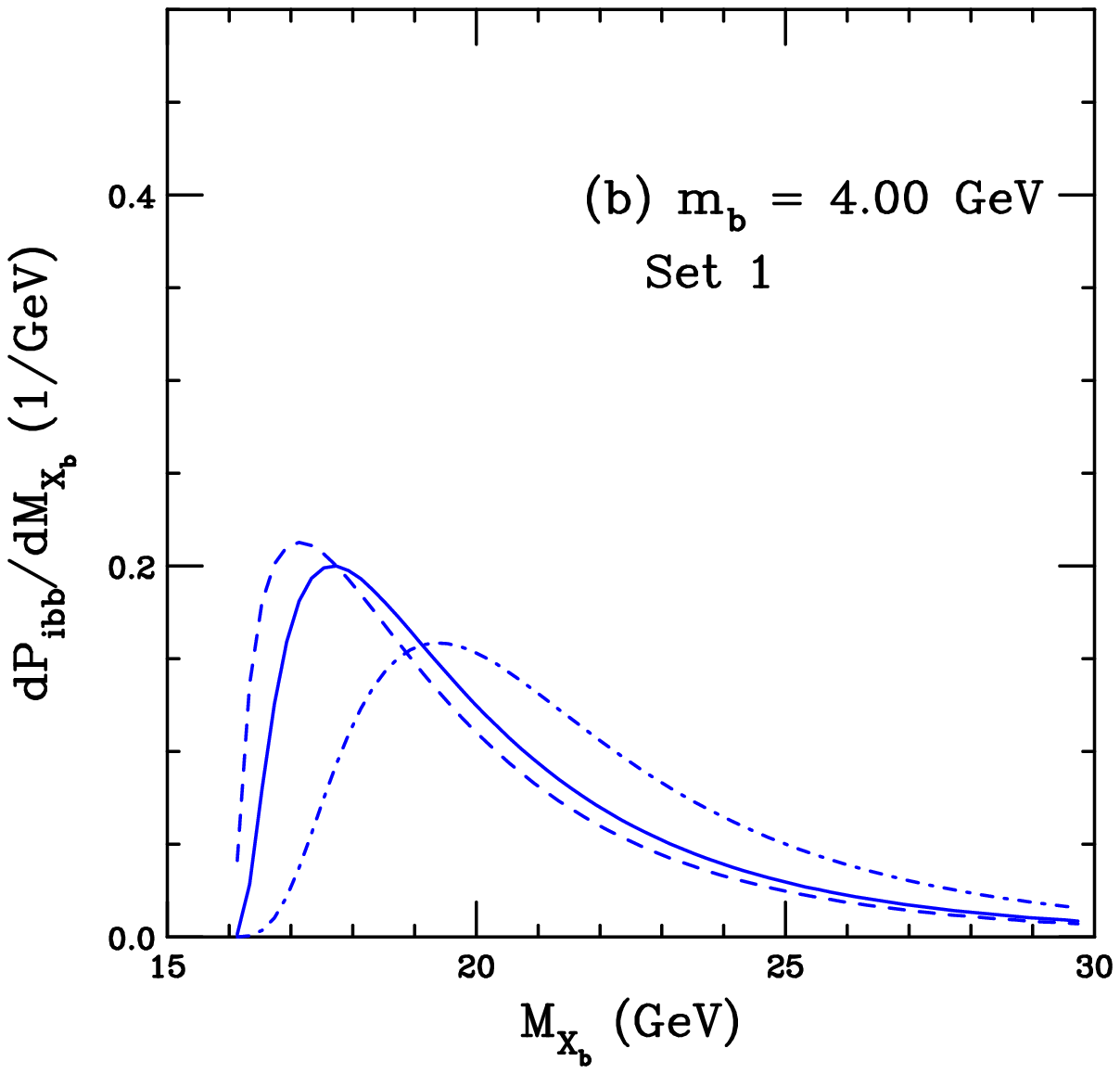}
  \end{center}
  \caption[]{(Color online) The probability for $X_b$ production
    from a seven-particle Fock state as a function of the $X_b$ mass for three
    different $k_T$ integration ranges, corresponding to Set 1 in
    Table~\protect\ref{ave_Xbmass_table}, are shown: $k_q^{\rm max} = 0.2$~GeV,
    different $k_T$ integration ranges are shown: $k_q^{\rm max} = 0.2$~GeV,
    $k_b^{\rm max} = 1.0$~GeV and $k_{X_b}^{\rm max} = 1.0$~GeV (solid);
    $k_q^{\rm max} = 0.1$~GeV, $k_b^{\rm max} = 0.5$~GeV and
    $k_{X_b}^{\rm max} = 0.5$~GeV (dashed); and
    $k_q^{\rm max} = 0.4$~GeV, $k_b^{\rm max} = 2.0$~GeV and
    $k_{X_b}^{\rm max} = 2.0$~GeV (dot-dashed).
    All distributions are normalized to unity.
    In (a) $m_b = 4.65$~GeV while in (b) $m_b = 4.0$~GeV.
  }
\label{ic_Xbdists}
\end{figure}

The mass distributions for the $X_b$ from Eq.~(\ref{Xb_mass}) are shown in
Fig.~\ref{ic_Xbdists} for the same values of $m_b$, $k_q^{\rm max}$ and
$k_b^{\rm max}$ as shown in Fig.~\ref{ic_Mdists} (Set 1 in
Table~\ref{ave_Xbmass_table}).  In this case the $k_T$ range
for the $X_b$ is taken to be the same as for the $b$ quarks.  The results here
are also normalized to unity.

The mass
distributions are much broader than those in Fig.~\ref{ic_Mdists}, even though
they have the same thresholds.  They become broader because there is no
integration over the $\Upsilon$ mass in this case, as in Eq.~(\ref{2Ups_mass}),
where the $\Upsilon$ mass was restricted to be between $2m_b$ and $2m_B$.
These broader distributions result in higher
average masses for the $X_b$, typically 1-2~GeV higher than for the same
Fock state where the $b \overline b b \overline b$ is considered to be bound
into an $\Upsilon$ pair.  Thus, no further systematic studies of the mass
dependence on $k^{\rm max}$ are
considered here.  
The average $X_b$ masses from these calculations
are given in Table~\ref{ave_Xbmass_table}, corresponding to
the first three rows of Table~\ref{ave_mass_table}.

\begin{table}
  \begin{center}
    \begin{tabular}{|c|c|c|c|c||c|c|c|c|} \hline
  Set &    \multicolumn{4}{|c||}{$m_b = 4.65$~GeV} & \multicolumn{4}{|c|}{$m_b = 4.00$~GeV} \\ \hline 
   &  $k_{q}^{\rm max}$ (GeV) & $k_{b}^{\rm max}$ (GeV) & $k_{X_b}^{\rm max}$ (GeV) & $\langle M_{X_b}\rangle$ (GeV) & 
$k_{q}^{\rm max}$ (GeV) & $k_{b}^{\rm max}$ (GeV) & $k_{X_b}^{\rm max}$ (GeV) & $\langle M_{X_b}\rangle$ (GeV)  \\ \hline 
  & 0.2 & 1.0 & 1.0 & 22.66 & 0.2 & 1.0 & 1.0 & 20.11 \\
1 & 0.1 & 0.5 & 0.5 & 22.31 & 0.1 & 0.5 & 0.5 & 19.65 \\
  & 0.4 & 2.0 & 2.0 & 23.72 & 0.4 & 2.0 & 2.0 & 21.47 \\ \hline
     \end{tabular}
  \end{center}
  \caption[]{The average $X_b$ mass for given values of the bottom quark mass and the maximum range of $k_T$ integration for light quarks, bottom quarks, and the $X_b$.}
  \label{ave_Xbmass_table}
\end{table}

It is worth noting that the probability for an $X_b$ state could be considerably
different from that of a double $\Upsilon$ state.  The factors of $F_B$ for
$\Upsilon$ production, as in Eq.~(\ref{sigib_UpsUps}), would be replaced by
a single factor $F_{X_b}$, giving
\be
\sigma_{{\rm ibb}\, 7}^{X_b} (pp) = F_{X_b} \, 
\frac{P_{{\rm ibb}\, 7}^0}{P_{{\rm ib}\, 5}^0}\  \sigma_{{\rm ib}\,5} (pp) \, \, .
\label{sigib_Xb}
\ee
In the CEM, $F_B$ is determined by a fit to $\Upsilon$ production data.  Since
no all bottom tetraquark states have so far been reported, aside from the
ANDY result which did not determine a cross section, the extraction of $F_{X_b}$
is not yet possible.

Finally, one could consider a tetraquark to be produced as a single massive
object rather than composed of two individual $b \overline b$ pairs.  In such a
case, the tetraquark
could be assumed to be produced in a four-particle Fock state,
$|uud X_b \rangle$, similar to a
massive gluon.  However, this configuration does not have a mass threshold and
would carry a lower share of the nucleon momentum on average,
$\langle x_{X_b} \rangle \sim 0.5$ as long as $\hat M_{X_b} \gg \hat m_q$.  Thus
this type of state would not conform to the kinematic criteria required for
either a double $\Upsilon$ state or an $X_b(b \overline b b\overline b)$ state.

\section{Summary}
\label{summary}

The ANDY Collaboration suggests that they have produced a
tetraquark $b \overline b b \overline b$ state with an average mass of
18.15~GeV \cite{ANDY}.  
The calculations for $\Upsilon$ pair and $X_b$ production from a seven-particle
double intrinsic bottom state presented here are incompatible with the low
mass suggested from the ANDY data.  However, the assumption of a double
$\Upsilon$ state produced from a seven-particle
$|uud b \overline b b\overline b\rangle$ state is compatible with previously
predicted tetraquark masses and could be considered a source of tetraquark
production via a double $\Upsilon$(1S) state. \\

{\bf Acknowledgments}
R.~V. would like to thank S.~J.~Brodsky and V.~Cheung for discussions.
This work was supported by the Office of Nuclear
Physics in the U.S. Department of Energy under Contract DE-AC52-07NA27344 and
the LLNL-LDRD Program under Project No. 21-LW-034.

\end{document}